\documentclass[pra,superscriptaddress,twocolumn,floatfix,showpacs,preprintnumbers,amsmath,amssymb,amsfonts,lengthcheck]{revtex4-2}

\usepackage[dvips]{graphics}

\DeclareMathAlphabet{\mathsfit}{\encodingdefault}{\sfdefault}{m}{sl}
\SetMathAlphabet{\mathsfit}{bold}{\encodingdefault}{\sfdefault}{bx}{sl}

\usepackage[dvips]{graphics}
\usepackage{geometry}
 \usepackage{ctable}
 \usepackage[percent]{overpic}

\usepackage{textgreek}
\usepackage{bm}
\usepackage{contour}
\usepackage{braket}
\usepackage{soul}

\usepackage{blindtext}
\usepackage[percent]{overpic}
\usepackage{mathtools}
\usepackage{xcolor}
\usepackage{multirow}
\usepackage{array}

\usepackage[percent]{overpic}

\usepackage{mathtools}
\renewcommand{\vec}[1]{\mathbf{#1}}

\renewcommand{\vec}[1]{\mathbf{#1}}

\newcommand{\tens}[1]{\mbox{\textsf{\textbf{#1}}}}

\newcommand{\Greektens}[1]{\contour[3]{black}{#1}}

\newcommand{\dif}{\!\! \mathrm{d}}
\newcommand{\mi}{\textrm{i}} 
\newcommand{\me}{\mathrm{e}}

\newcommand{\Et}{\vec{E}(\vec{r},t)}

\newcommand{\Eclo}{\mathcal{E}^{(1)}}
\newcommand{\Eclt}{\mathcal{E}^{(2)}}
\newcommand{\Ecli}{\mathcal{E}^{(i)}}
\newcommand{\Ei}{\hat{E}^{(i)}}
\newcommand{\Eo}{\hat{E}^{(1)}}
\renewcommand{\Et}{\hat{E}^{(2)}}

\begin{document}

\title{Probing Vacuum Field Fluctuations and Source Radiation Separately in Space and Time}

\author{Frieder Lindel}
\affiliation{\freiburg}
\author{Alexa Herter}
\affiliation{\ethZ}
\author{J\'{e}r\^{o}me Faist}
\affiliation{\ethZ}
\author{Stefan Yoshi Buhmann}
\affiliation{\kassel}

\newcommand{\freiburg}{Physikalisches Institut, Albert-Ludwigs-Universit\"{a}t Freiburg, Hermann-Herder-Stra{\ss}e 3, D-79104, Freiburg, Germany}
\newcommand{\kassel}{Institut f\"{u}r Physik, Universit\"{a}t Kassel, Heinrich-Plett-Stra{\ss}e 40, 34132 Kassel, Germany}
\newcommand{\ethZ}{ETH Zurich, Institute of Quantum Electronics, Auguste-Piccard-Hof 1, 8093 Zurich, Switzerland}

\date{\today}

\date{\today}

\begin{abstract}
Source radiation (radiation reaction) and vacuum-field fluctuations can be seen as two inseparable contributions to processes such as spontaneous emission, the Lamb shift, or the Casimir force. Here, we propose how they can be \textit{individually} probed and their space-time structure revealed in electro-optic sampling experiments. This allows to experimentally study causality at the single photon level and to reveal space- and time-like correlations in the quantum vacuum. A connection to the time-domain fluctuation--dissipation theorem is also made.  
\end{abstract}

\maketitle

\section{Introduction}

Analyzing the physical mechanisms behind the structure and dynamical properties of an atom interacting with the electromagnetic vacuum, e.g., the Lamb shift and spontaneous emission, one finds two inseparable contributions \cite{milonni1994quantum}: (i) Fluctuations of the atom's charged constituents lead to the emission of source radiation which can act back on the atom (radiation reaction) \cite{welton_observable_1948,ackerhalt_radiation_1973-1,milonni_radiation_1981,milonni_casimir_1982}, (ii) the atom can interact with vacuum fluctuations of the electromagnetic field \cite{welton_observable_1948}. It is the intricate interplay between these two contributions which leads to the stability of ground-state atoms: The loss of energy due to the emission of source radiation by the fluctuating charges is canceled by the process in which the atom absorbs energy from the vacuum \cite{milonni1994quantum}.

\begin{figure*}
\includegraphics[width=2.\columnwidth]{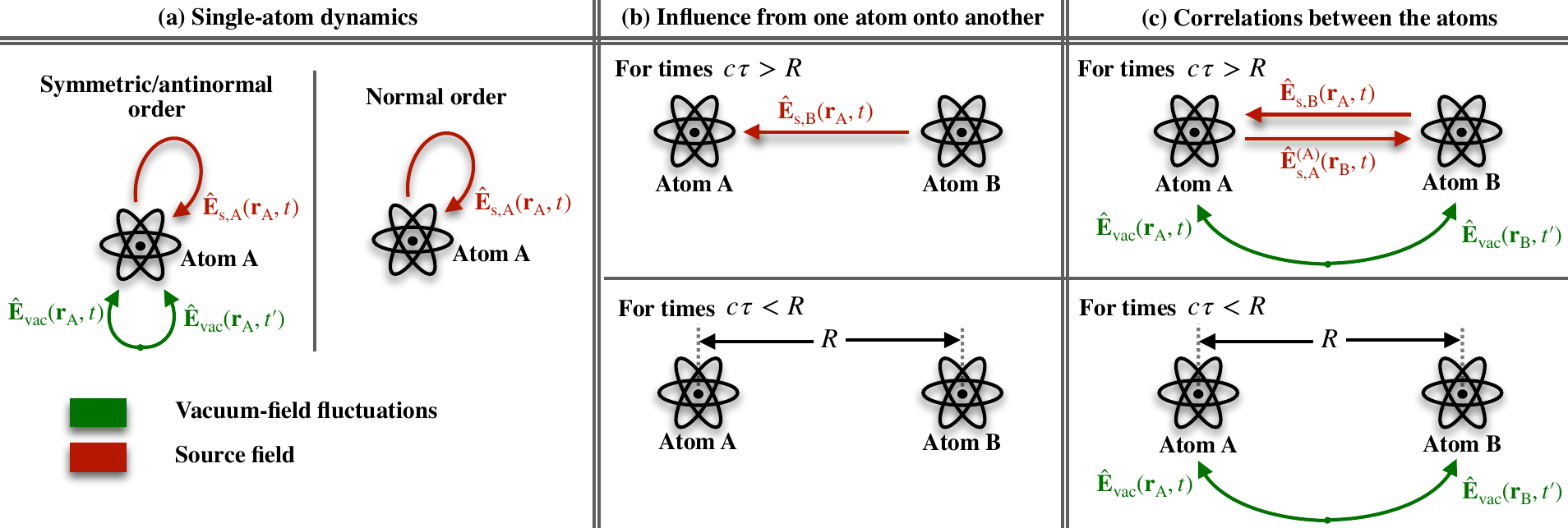}
\caption{\textit{Source radiation vs.~vacuum-field fluctuations for single atoms dynamics or the Fermi two-atom setup.} (a) The structure and dynamical properties of a single atom coupled to the electromagnetic field in vacuum is influenced by vacuum-field fluctuations (red) and radiation reaction (green), i.e., the back action of the source field emitted by the atom onto itself. Although the sum of these effects remains independent of the chosen operator orderings, their individual contributions differ, e.g., the vacuum field contribution to spontaneous emission vanishes using normal ordering while it does not vanish for symmetric or antinormal ordering. (b, c) Fermi's two-atom setup: Two atoms at a distance $R$ apart from each other interact for a finite time $\tau$ with the electromagnetic vacuum field. (b) The influence of atom $B$ onto atom $A$ is always (independent on the chosen operator ordering) described by the fully retarded source radiation of atom $B$, $\hat{E}_\mathrm{s,B}$. As $\hat{E}_\mathrm{s,B}$ vanishes for space-like separated atoms, there is no faster-than-light signaling. (c) Correlations between the two atoms arise either due to an exchange of source radiation or since correlations from the vacuum field are swapped to the atoms. Only for symmetric operator ordering (as in the DDC formalism) the former contribution vanishes for space-like separated atoms, as illustrated above. }
\label{fig:AtomsScheme}
\end{figure*}

These `two sides of the same quantum-mechanical coin' \cite{senitzky_radiation-reaction_1973} also play the pivotal roles in Fermi's two-atom Gedankenexperiment \cite{fermi_quantum_1932,tjoa_when_2021}, in which two atoms are placed at a distance $R$ in empty space and interact with the vacuum electromagnetic field for a finite time $\tau$, see Figs.~\ref{fig:AtomsScheme} (b) and (c). When considering the generation of correlations between the two atoms \cite{biswas_virtual_1990} one finds that the atoms become correlated since atom A interacts with the source radiation emitted by atom B (or vice versa), i.e., the two atoms exchange photons, or they can individually interact with vacuum-field fluctuations and thereby correlations pre-existing in the vacuum field are swapped to the atoms  \cite{tjoa_when_2021}.

The influence of source radiation and vacuum-field fluctuations onto the dynamics of atoms, however, can not be uniquely identified in general, since their relative contributions depend on the chosen initial operator ordering in the Hamiltonian \cite{senitzky_radiation-reaction_1973,milonni_interpretation_1973}, see Fig.~\ref{fig:AtomsScheme} (a). This indetermination can be removed by the requirements that (i) the source and vacuum-field contributions must be individually hermitian and (ii) expressed in terms of hermitian field and atomic operators. This is known as the DDC (Dalibard, Dupont-Roc, Cohen-Tannoudji) procedure \cite{dalibard_vacuum_1982,dalibard_dynamics_1984}.

Here, in Sec~\ref{sec:FermiTwoAtom}, we revisit Fermi's two-atom setup and show that to lift the indetermination of the relative contributions of source radiation and vacuum-field fluctuations one can replace (ii) in the DDC procedure by the following equivalent causality requirement: the source field should not lead to correlations between space-like separated atoms. This leads to the same unique operator ordering as in the DDC formalism and implies that any other operator ordering suffers from the complication that either the source and vacuum-field contributions are in general not real numbers or the source field mediates correlations faster than the speed of light.

The interpretation of the resulting unique source and vacuum-field contributions are in-line with previous results, discussing how fundamental features of quantum field theory can be revealed in Fermi's two atom setup: \textit{(a) Existence of space- and time-like correlations in vacuum} \cite{summers_vacuum_1985,summers_bells_1987,summers_bells_1987-1,franson_generation_nodate,biswas_virtual_1990,olson_entanglement_2011}: Space-like separated atoms, so atoms which are completely causally disconnected from each other, can become correlated, only because they can harvest space-like correlations existing in the quantum vacuum \cite{biswas_virtual_1990,reznik_violating_2005,summers_vacuum_1985}. As no information is transferred between the atoms via this process, this result is in accordance with special relativity \cite{biswas_virtual_1990}. Similarly, if the separation of the two atoms is purely time-like, then they also cannot exchange a photon traveling at the speed of light. Still they can become correlated by extracting so-called past--future correlations from the quantum vacuum \cite{olson_entanglement_2011,olson_extraction_2012}. \textit{(b) Causality:} The two atoms can infer information about each other via the exchange of source radiation. According to special relativity, its contribution must therefore vanish, when the atoms remain space-like separated ($R > c \tau $, $c$: speed of light). Whether this is strictly true in a quantum mechanical treatment, had led to a debate in the past \cite{fermi_quantum_1932,biswas_virtual_1990,buchholz_there_1994,hegerfeldt_causality_1994}, which was eventually settled theoretically in favor of strict causality \cite{milonni_photodetection_1995,power1997analysis}. An experimental verification \cite{sabin_fermi_2011-1} is, however, still missing. \textit{(c) Time-domain fluctuation--dissipation theorem (FDT)} \cite{pottier_quantum_2001}: The FDT is usually considered in frequency space. Here, we find a direct implication of its time-domain form: It relates correlations arising from source radiation or vacuum-field fluctuations in Fermi's two-atom problem. This also allows for an experimental verification of the time-domain FDT.   

The experimental implementation \cite{zohar_fermi_2011,sabin_fermi_2011-1} of the Fermi two-atom setup is challenging, especially since the interaction with the vacuum has to be switched on and off on very short time-scales. Recently, an analog version has been introduced \cite{settembrini2022detection} based on electro-optic sampling (EOS) \cite{wu1995free,wu1996ultrafast,riek_direct_2015,moskalenko_paraxial_2015,benea-chelmus_electric_2019,virally2021enhanced,onoe2022realizing,hubenschmid2022complete,gundougdu2023self}, leading to the first observation of space-like correlations in the quantum vacuum \cite{settembrini2022detection}: 
Two tightly focused laser pulses, separated in space and time by $\delta r$ and $\delta t$, respectively, propagate through a nonlinear crystal. They effectively induce an interaction between the THz quantum-vacuum field and two near-infrared (NIR) field modes (replacing the atoms) via the nonlinear coupling of the crystal (laser pulses entering and leaving the crystal switches the interaction on and off), see Fig.~\ref{fig:TwoLevelSystem}(b).

In Sec \ref{sec:EOS}, we analyze the potential of EOS experiments and find that source-radiation and vacuum-field contributions can be accessed \textit{individually}. This is achieved by simply exchanging two wave-plates in the detection scheme. We then show how this allows one to probe $(a)$--$(c)$ experimentally with state-of-the-art EOS setups.

\section{Vacuum and Source Fields in Fermi's Two-Atom Setup} \label{sec:FermiTwoAtom}

We consider Fermi's two-atom setup \cite{fermi_quantum_1932} consisting of two two-level atoms interacting for a finite time $\tau$ with the quantized electromagnetic field initially in its vacuum state. The interaction Hamiltonian reads 
\begin{align} \label{eq:IAHamiltonianFermi}
\hat{H}_I^{(F)}(t)= - \sum_{i=A,B} \eta(t)  \int \dif^3 r F(\vec{r}- \vec{r}_i)\hat{E}(\vec{r},t) \hat{d}^{(i)}(t).
\end{align} 
Here, $\hat{d}^{(A)}$ and $\hat{d}^{(B)}$ are the dipole operators of atom $A$ and $B$, respectively, $\hat{E}$ is the electric field operator, $r_A$ and $r_B$ are the center of mass positions of the smeared out atoms with smearing function $F$, and $\eta(t)$ is a switching function, which switches the interaction between field and atoms on and off on a time scale $\tau$. For simplicity, we assumed a unidirectional dipole moment of the two atoms, requiring consideration of only one field polarization direction.

Note, that all conclusions of this section qualitatively also apply to the more generic setting, in which two quantum systems (here the atoms) are interacting linearly and weakly with a joint bath (here the electromagnetic field), cf. Refs.~\cite{dalibard_vacuum_1982,dalibard_dynamics_1984}.

\subsection{Source and Vacuum-Field Contributions to Single-Atom Dynamics and Causality}

We follow Refs.~\cite{dalibard_vacuum_1982,milonni1994quantum} to identify source and vacuum-field contributions to single-atom dynamics. As the interaction Hamiltonian in Eq.~\eqref{eq:IAHamiltonianFermi} is linear in the electric field operator, the solution to Heisenberg's equation of motion for the field is given by \cite{milonni1994quantum}
\begin{align} \label{eq:FermiEFieldSolution}
\hat{E}(\vec{r},t)= \hat{E}_\text{vac}(\vec{r},t) + \hat{E}_\text{s,A}(\vec{r},t)+\hat{E}_\text{s,B}(\vec{r},t),
\end{align}
where $ \hat{E}_\text{vac}$ is just the vacuum electric field in absence of the two atoms, and $ \hat{E}_{\text{s},i}$ is the source radiation emitted by atom $i$. 

To identify the influence of source radiation and vacuum-field fluctuations onto the atoms' dynamics, we insert the field in Eq.~\eqref{eq:FermiEFieldSolution} into the equation of motion for the atomic observable of interest $\hat{O}$. The terms proportional to $\hat{E}_\text{vac}(\vec{r},t) $ and to $\hat{E}_\text{s,A}(\vec{r},t)+\hat{E}_\text{s,B}(\vec{r},t)$ are then identified as the vacuum and source field contributions, respectively, see Ref.~\cite{dalibard_vacuum_1982,dalibard_dynamics_1984} and also Appendix \ref{app:FermiSingle}. In the former contribution, the atoms respond to fluctuations in the field, while in the latter, fluctuations of the atomic dipoles induce a response in the field, which acts back on the atoms.

In the equation of motion of $\hat{O}$, the total electric field $\hat{E}$ and $\hat{d}^{(i)}$ commute, such that we can write them in any order. The overall dynamics are, of course, independent of the chosen operator ordering. The vacuum and source-field contributions, however, individually depend on the chosen operator ordering \cite{senitzky_radiation-reaction_1973,milonni_interpretation_1973,dalibard_vacuum_1982}, since $\hat{E}_\mathrm{vac}$ and $\hat{E}_{\mathrm{s},i}$ do not commute with $\hat{d}^{(i)}$, respectively. For example, it has been shown that using normal operator ordering (all positive [negative] frequency components of the field ordered to the right [left]) spontaneous emission arises only due to source-radiation (radiation reaction), whereas using symmetric operator ordering vacuum-field effects also contribute \cite{welton_observable_1948,milonni1994quantum}, see Fig.~\ref{fig:AtomsScheme} (a) and (b). We thus find an indetermination in the separation into source and vacuum field contributions, which can only be lifted by an additional constraint. Refs.~\cite{dalibard_vacuum_1982,dalibard_dynamics_1984} suggest to use a symmetric ordering, as only in this case source and vacuum-field contributions remain individually hermitian and are expressed in terms of hermitian field and atomic operators. This procedure is known as the DDC formalism, which has been used to identify source and vacuum-field contributions for a variety of different processes, see, e.g., Refs.~\cite{milonni1994quantum,dalibard_vacuum_1982,dalibard_dynamics_1984,rizzuto_casimir-polder_2007,rizzuto_lamb_2009,rizzuto_energy-level_2011,menezes_radiative_2016,marino_thermal_2014,cheng2023general,zhou_radiation-reaction-induced_2020}. 

This indetermination is not present when it comes to the question of causality in Fermi's two-atom problem, i.e., whether faster-than-light signaling is possible, see Fig.~\ref{fig:AtomsScheme} (b). Atom $A$ only interacts with atom $B$ via the field in Eq.~\eqref{eq:FermiEFieldSolution}. Thus, atom $A$ can only infer the presence of atom $B$ by interacting with $\hat{E}_\mathrm{s,B}$. As we show in Appendix \ref{app:FermiSingle}, this contribution is independent of the chosen operator ordering. As pointed out in Ref.~\cite{milonni_photodetection_1995,milonni1994quantum}, $\hat{E}_{\mathrm{s},i}$ is fully retarded, which ensures that no-faster-then-light signaling is possible in Fermi's two atom setup.

\subsection{Correlations} \label{sec:FermiTwoAtomCorr}

We quantify correlations between an observable of atom $A$ and atom $B$, $\hat{O}^{(A)}$ and $\hat{O}^{(B)}$, respectively, via the correlation function
\begin{align} \label{eq:FermiCorrelationFunction}
G^{(AB)}(t)  = \langle \hat{O}^{(A)} \hat{O}^{(B)}\rangle -\langle \hat{O}^{(A)}\rangle \langle \hat{O}^{(B)}\rangle.
\end{align}
Analogously to the case of single-atom observables discussed in the last section, we find that again both `sides of the same quantum-mechanical coin' \cite{senitzky_radiation-reaction_1973}, source radiation and the vacuum field, contribute to the dynamics of the correlation function in Eq.~\eqref{eq:FermiCorrelationFunction}. So far, to the best of our knowledge, for Fermi's two-atom setup these contributions have only been identified in Ref.~\cite{tjoa_when_2021}. However, in the interaction picture calculation employed in Ref.~\cite{tjoa_when_2021}, the connection to the above results on single-atom observables, including the issue of the initial operator ordering, could not be clarified.

Identifying source and vacuum-field contributions to the dynamics of the correlation function in Eq.~\eqref{eq:FermiCorrelationFunction} follows along similar lines as for the single-atom observables and is described in detail in Appendix \ref{app:FermiCorr}. The resulting expression for the source and vacuum field contributions again depend on the operator ordering. For symmetric ordering, as used in the DDC procedure, and assuming that the atoms are initially uncorrelated such that $G^{(AB)}(t_0)=0$, we find the following vacuum and source-field contributions to the two-point correlation function:
\begin{multline} \label{eq:SymmOrdVF}
G^{(AB)}(t)\big|_\mathrm{vac} = \int \dif^3 r^{\prime\prime} \int_{t_0}^t \dif t^{\prime\prime} \int \dif^3 r^\prime \int_{t_0}^t \dif t^\prime  \\
\times  L^{(A)}(\vec{r}^\prime, t^\prime, t) L^{(B)}(\vec{r}^{\prime\prime}, t^{\prime\prime}, t)\mathcal{C}(\vec{r}^\prime, \vec{r}^{\prime\prime} ,t^\prime-t^{\prime\prime}) ,
\end{multline}
and 
\begin{multline} \label{eq:SymmOrdSR}
G^{(AB)}(t)\big|_\mathrm{s} =\int \dif^3 r^{\prime\prime} \int_{t_0}^t \dif t^{\prime\prime} \int \dif^3 r^\prime \int_{t_0}^t \dif t^\prime  \\
\times  L^{(A)}(\vec{r}^\prime, t^\prime, t) \tilde{L}^{(B)}(\vec{r}^{\prime\prime}, t^{\prime\prime}, t)\mathcal{R}(\vec{r}^\prime, \vec{r}^{\prime\prime} ,t^\prime-t^{\prime\prime})  + A \leftrightarrow B.
\end{multline}
Here, $+ A \leftrightarrow B$ means adding the previous term subject to the replacement $A \leftrightarrow B$, and we defined 
\begin{align}
L^{(i)}(\vec{r}^\prime, t^\prime, t) =\frac{\mi}{\hbar} \eta(t^\prime) F(\Delta \vec{r}^\prime_i) \braket{[\hat{d}^{(i)}_\mathrm{vac}(t^\prime ),\hat{O}^{(i)}_\mathrm{vac}(t )] } ,
\end{align}
and
\begin{multline}
\tilde{L}^{(i)}(\vec{r}^\prime, t^\prime, t) =\frac{1}{2} \eta(t^\prime) F(\Delta \vec{r}^\prime_i) \big( \braket{\{\hat{d}^{(i)}_\mathrm{vac}(t^\prime ),\hat{O}^{(i)}_\mathrm{vac}(t )\} }    \\ -2 \braket{ \hat{d}^{(i)}_\mathrm{vac}(t^\prime )}\braket{ \hat{O}^{(i)}_\mathrm{vac}(t ) } \big) ,
\end{multline}
with $ \Delta  \vec{r}^\prime_i = \vec{r}^\prime- \vec{r}_i$, and $\{\cdot, \cdot\}$ denotes the anticommutator. The electric field operator enters Eq.~\eqref{eq:SymmOrdVF} and \eqref{eq:SymmOrdSR} via its correlation and response function, which are given by
\begin{align} \label{eq:CorrelationFunction}
\mathcal{C}(\Greektens{$\rho$}, \tau) & = \frac{1}{2}\langle \{ \hat{E}_\mathrm{vac}(\vec{r}, t ), \hat{E}_\mathrm{vac}(\vec{r}^\prime, t^\prime) \} \rangle, \\ \label{eq:ResponseFunction}
\mathcal{R}(\Greektens{$\rho$} ,\tau)  & = \frac{\mi }{\hbar}  \theta(\tau)  [\hat{E}_{\mathrm{vac}}(\vec{r}, t) , \hat{E}_{ \mathrm{vac}}(\vec{r}^\prime, t^\prime) ] .
\end{align}
Here, \mbox{$\tau = t-t^\prime$}, $\Greektens{$\rho$} = \vec{r}-\vec{r}^\prime$, and $\theta(\tau)$ is the Heaviside step function. $\mathcal{C}$ and $\mathcal{R}$ are well known quantities from linear response theory, see Appendix \ref{app:Prelim} for a brief summary of their main characteristics. In free space they read 
\begin{align}  \label{eq:CorrelationFree}
\mathcal{C}(\Greektens{$\rho$}, \tau) & = \frac{\mu_0 \hbar}{8 \pi^2   } \square \frac{1}{\rho} \left(\frac{\mathcal{P}}{\frac{\rho}{c}- \tau} + \frac{\mathcal{P}}{\frac{\rho}{c}+ \tau}\right), \\  \label{eq:ResponseFree}
\mathcal{R}(\Greektens{$\rho$}, \tau) & = \frac{ \mu_0}{4 \pi   } \square \frac{1}{\rho}  \delta\left(\frac{\rho}{c}-\tau \right) , 
\end{align}
with $\square =\frac{\partial^2}{\partial t \partial t^\prime} - c^2 \frac{\partial^2}{\partial x \partial x^\prime}$ and $\mathcal{P}$ the Cauchy principal value. We see from Eqs.~\eqref{eq:CorrelationFree} and \eqref{eq:ResponseFree} that the response function $\mathcal{R}$ only has support along the light-cone, i.e., if $\rho \equiv |\vec{r}-\vec{r}^\prime| = c \tau$, whereas the correlation function does not vanish for space- or time-like separations. Equations~\eqref{eq:SymmOrdVF} and \eqref{eq:SymmOrdSR} thus give the expected result, which is in-line with the discussion of Fermi's two-atom setup found in previous works \cite{biswas_virtual_1990,franson_generation_nodate,milonni_photodetection_1995,tjoa_when_2021}: Either the two atoms get correlated due to an exchange of source radiation, propagating at the speed of light from one atom to the other (source-field contribution), or they \textit{individually} interact with the vacuum field and thereby correlations in the vacuum field are swapped to the atoms (vacuum-field contribution), see Fig.~\ref{fig:AtomsScheme} (c). Thus, correlations between space- or time-like separated atoms can only arise due to vacuum-field contributions. 

The particular separation into source-radiation and vacuum-field contributions in Eqs.~\eqref{eq:SymmOrdVF} and \eqref{eq:SymmOrdSR}, however, is only found using symmetric operator ordering, see Appendix \ref{app:FermiCorr}. For all other orderings, either the source-radiation contribution leads to correlations between space- and time-like separated atoms, or the source and vacuum-field contributions are not given by real numbers, see Appendix \ref{app:FermiCorr}. For example, in case of normal ordering we find that the vacuum-field contribution vanishes, i.e., $G^{(AB)}(t)|_\mathrm{vac} = 0$ whereas $G^{(AB)}(t)|_\mathrm{s}$ is given by the sum of the right-hand sides of Eqs.~\eqref{eq:SymmOrdVF} and \eqref{eq:SymmOrdSR}. 

To conclude this discussion, for single atom observables, the DDC procedure is motivated by the requirements that (i) the individual contributions from the source and vacuum field should be individually hermitian and (ii) expressed in terms of hermitian field and atomic operators. Considering correlations in the Fermi two-atom setup, one finds that (ii) can also be replaced by the physical requirement that the source radiation contribution does not induce correlations between space- or time-like separated emitters, allowing one to interpret this contribution as mediated by a field propagating at the speed of light from one atom to the other.

Interestingly, using the symmetric operator ordering, there are certain choices of $\hat{O}^{(A)}$ and $\hat{O}^{(B)}$ for which the source-radiation contribution vanishes while the vacuum-field one does not, and vice versa: We find that $G^{(AB)}(t)|_\mathrm{vac} = 0$ and $G^{(AB)}(t)|_\mathrm{s} \neq 0$ if $\hat{O}^{(A)} = \hat{\sigma}_x^{(A)}$ and $\hat{O}^{(B)} = \hat{\sigma}_y^{(B)}$, while  $G^{(AB)}(t)|_\mathrm{vac} \neq 0$ and $G^{(AB)}(t)|_\mathrm{s} = 0$ if $\hat{O}^{(A)} = \hat{\sigma}_y^{(A)}$, $\hat{O}^{(B)} = \hat{\sigma}_y^{(B)}$, and each atom is initially either in the ground or the excited state ($\sigma_x$ and $\sigma_y$ are Pauli matrices, see Appendix \ref{app:Fermi}). This implies that only source radiation leads to correlations between $\hat{\sigma}_x^{(A)}$ and $\hat{\sigma}_y^{(B)}$, while only the vacuum field correlates $\hat{\sigma}_y^{(A)}$ and $\hat{\sigma}_y^{(B)}$. This allows one to access these two `inseparable two sides of the same coin' \cite{senitzky_radiation-reaction_1973}, i.e., source radiation and vacuum field effects, \textit{individually} by probing correlations between specific single-atom observables. Changing the time interval and points in space at which the atoms interact with the electromagnetic field further allows one to individually probe the space-time structure of these two contributions \cite{tjoa_when_2021}. We analyze how this can be experimentally achieved in the EOS analog of Fermi's two-atom setup in Sec.~\ref{sec:EOS}. 

Note that the correlation and response function of the field in Eqs.~\eqref{eq:CorrelationFree} and \eqref{eq:ResponseFree} are connected via the fluctuation--dissipation theorem. This also connects the source and vacuum-field contributions in Eqs.~\eqref{eq:SymmOrdVF} and \eqref{eq:SymmOrdSR}, which is again further discussed in Sec.~\ref{sec:FDT} for the EOS analog of Fermi's two-atom setup.

\section{Electro-Optic Sampling} \label{sec:EOS}

In the following, we analyze how source radiation and vacuum-field fluctuations lead to correlations in the two-beam EOS setup, which has been experimentally realized in Refs.~\cite{benea-chelmus_electric_2019,settembrini2022detection}, and which can be seen as an analog of Fermi's two-atom setup discussed in the last section \cite{settembrini2022detection}.

\subsection{Experimental Setup }

\begin{figure}
\includegraphics[width=1.\columnwidth]{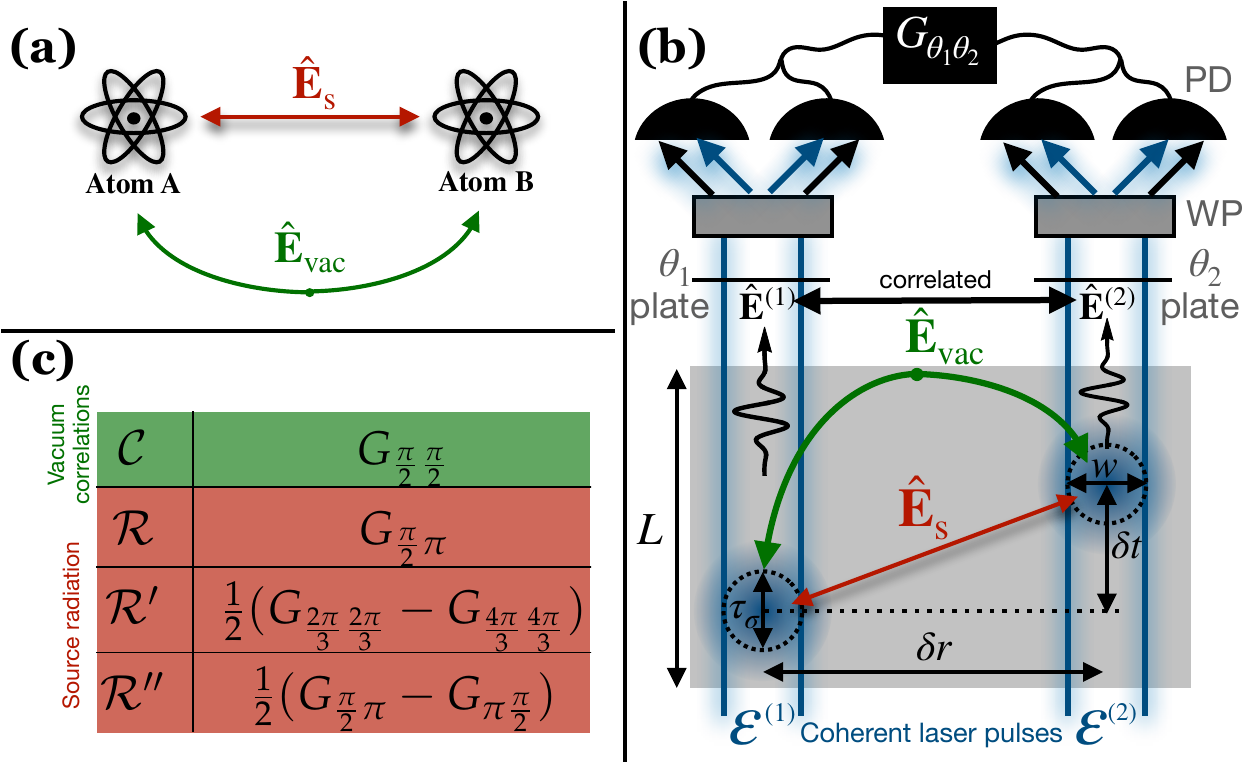}
\caption{\textit{EOS analog of Fermi's two-atom setup.} (a) Two atoms interacting with the quantized electromagnetic field in its vacuum state can either become correlated by exchanging source radiation or by `harvesting' correlations existing in the vacuum field $\hat{E}_\mathrm{vac}$, see also Fig.~\ref{fig:AtomsScheme}. (b) EOS setup: Analogously, correlations between two probe-field modes $\hat{E}^{(1,2)}$ can arise, which co-propagate with two coherent laser pulses $\mathcal{E}^{(1,2)}$ through a nonlinear crystal (gray), in which they can effectively interact with the THz electric field $\hat{E}$. These correlations can be accessed via a balanced detection scheme for each laser pulse consisting of a wave-plate with angles $\theta_{1,2}$, a Wollaston prism (WP) and two balanced photo-diodes (PD). 
(c) For certain $\theta_{1,2}$ the EOS signal $G_{\theta_1\theta_2}$ is only sensitive to correlations stemming from source radiation or vacuum field fluctuations, given by the response ($\mathcal{R}$) and correlation ($\mathcal{C}$) function, respectively.}
\label{fig:TwoLevelSystem}
\end{figure}

We consider the two-beam EOS setup in Fig.~\ref{fig:TwoLevelSystem} (b): Two tightly-focused, uncorrelated, near-infrared coherent probe pulses $\mathcal{E}^{(1)} $ and $\mathcal{E}^{(2)}$, linearly polarized into $y$ direction, are propagating along the $[110]$-axis ($z$-axis) through a zinc-blende type nonlinear crystal. Inside the crystal they can effectively interact via the nonlinear coupling of the crystal with the $x$-polarized THz quantum field $\hat{E} $ (parallel to $[\overline{1}10]$), which is initially in its vacuum state $\hat{E}_\mathrm{vac} $ \cite{moskalenko_paraxial_2015,lindel2020theory,lindel2021macroscopic}. The interaction is given by the Hamiltonian \cite{onoe2022realizing}
\begin{align} \label{eq:HiEOS}
\hat{H}_{I}(t) = 2 \chi^{(2)} \sum_{i = 1,2} \int_{V_C} \dif ^3 r  \mathcal{E}^{(i)}(\vec{r}, t) \hat{E} (\vec{r}, t)  \hat{E}^{(i)} (\vec{r}, t),
\end{align}
 where $V_C$ is the crystal volume and $ \chi^{(2)}$ the nonlinear susceptibility. The interaction of the laser pulses with the quantum field $\hat{E} $ leads to $x$-polarized NIR signal fields $\hat{E}^{(1)} $, $\hat{E}^{(2)} $ emerging from the crystal, which are co-propagating with the coherent fields $\mathcal{E}^{(1)} $, $\mathcal{E}^{(2)} $, respectively, due to phase-matching constraints. 
 
Comparing Eqs.~\eqref{eq:IAHamiltonianFermi} and \eqref{eq:HiEOS}, we see the close similarity between the two-beam EOS and Fermi's two-atom setup, as has been already discussed in Refs.~\cite{onoe2022realizing,settembrini2022detection}. In Fermi's two-atom problem, two two-level systems are coupled to the electromagnetic field $\hat{E}$ via their dipoles. The effective space-time volume in which the interaction takes place, is determined by their spatial smearing and the switching function $F$ and $\eta$, respectively. In the EOS analog, the two atoms are replaced by two paraxial field modes $\hat{E}^{(i)}$, which are co-propagating with the focused laser pulse $\mathcal{E}^{(i)}$ and couple to the vacuum field via the nonlinear coupling in the crystal. Here, the interaction takes place in the space-time volume of the ultrashort laser pulses inside the crystal given by $\int_{V_C} \dif ^3 r  \mathcal{E}^{(i)}(\vec{r}, t) $ in Eq.~\eqref{eq:HiEOS}. Note, that $\hat{E}^{(1)} $ and $\hat{E}^{(2)} $ are initially in their vacuum state, corresponding to two ground-state atoms in the Fermi two-atom setup. As we show below, in close analogy to the atoms in Fermi's two-atom setup, $\hat{E}^{(1)} $ and $\hat{E}^{(2)} $ become correlated either because they exchange source radiation or because correlations from the quantum-vacuum field $\hat{E}_\mathrm{vac} $ are swapped to $\hat{E}^{(1)} $ and $\hat{E}^{(2)} $. To measure these correlations we consider the EOS signal $G_{\theta_1 \theta_2}$ which probes correlations between the signal fields $\hat{E}^{(1)} $, $\hat{E}^{(2)} $ via the correlations between two ellipsommetry measurements $\hat{S}^{(1)}(\theta_1)$ and $ \hat{S}^{(2)}(\theta_2)$ of the fields emerging from the crystal, i.e., \cite{Note5}
\begin{multline} \label{eq:SignalG}
 G_{\theta_1 \theta_2} = \frac{1}{C} \big[ \langle \hat{S}^{(1)}(\theta_1) \hat{S}^{(2)}(\theta_2)   \rangle \\-\langle \hat{S}^{(1)}(\theta_1) \rangle \langle \hat{S}^{(2)}(\theta_2)   \rangle \big] ,
\end{multline}
   with the detection efficiency $C$ defined in Eq.~\eqref{eq:Detector}, and \cite{moskalenko_paraxial_2015,kizmann_quantum_2022}:
\begin{multline} \label{eq:Signal}
\hat{S}^{(i)}(\theta_i) = 4\pi \epsilon_0 c n_\mathrm{c} \int \dif^2 r_\parallel \int_0^\infty \dif \omega \frac{1}{\hbar \omega} \\\times [P(\theta_i) \mathcal{E}^{(i)\ast}(\vec{r}_\parallel, \omega)  \hat{E}^{(i)}(\vec{r}_\parallel, \omega) + \mathrm{h.c.}].
\end{multline}
Here, $\epsilon_0$ is the vacuum permittivity and $n_\mathrm{c}$ the refractive index at the central frequency of the laser pulses. $P(\theta_i) = \sqrt{-\mathrm{cos}(\theta_i)} + \mi \sqrt{2} \mathrm{cos} (\theta_i/2)$ accounts for the influence of wave plates in the detection setup, which induce phase shifts $\theta_{i} \in [\pi/2, 3 \pi/2]$ between the signal fields ($\hat{E}^{(i)}$) and the laser pulses ($\mathcal{E}^{(i)}$) after emerging from the crystal, see Fig.~\ref{fig:TwoLevelSystem}(b). Such a detection scheme was recently considered in the single-beam EOS setup \cite{sulzer2020determination,onoe2022realizing,kizmann_quantum_2022,hubenschmid2022complete}, whereas previous experimental \cite{benea-chelmus_electric_2019,settembrini2022detection} and theoretical \cite{lindel2020theory,lindel2021macroscopic} works on the two-beam EOS setup only considered $\theta_{1,2} = \pi/2$. As we will discuss below, tuning the two different angles $\theta_{1,2}$ allows one to individually probe contributions from vacuum fluctuations and source radiation.

To obtain the EOS signal, we solve Heisenbergs equations of motion for the quantum and signal fields $\hat{E}$ and $\hat{E}^{(i)}$ perturbatively in orders of $\chi^{(2)}$. This follows along the same lines as the calculation of the correlations in Fermi's two atoms setup discussed in Sec.~\ref{sec:FermiTwoAtomCorr}, see Appendix \ref{app:EOS} for details. Employing a symmetric operator ordering as motivated in Sec.~\ref{sec:FermiTwoAtom}, we find that up to second order in $\chi^{(2)}$ there are two different processes which lead to correlations between $\hat{E}^{(1)}$ and $\hat{E}^{(2)}$, and, thus, to an EOS signal $G_{\theta_1\theta_2}$ [see Fig.~\ref{fig:TwoLevelSystem}(b)]: Either the two laser pulses individually mix with the vacuum fluctuations $\hat{E}_\mathrm{vac}$, thereby harvesting correlations from the latter, or one of the laser pulses generates THz source radiation $\hat{E}_s$ which propagates to and then interacts via the nonlinear coupling with the other laser pulse.

\subsection{Vacuum-Field Contribution}

We start discussing the former process. The mixing of the laser pulses with the vacuum field via one nonlinear process (sum- or difference-frequency generation processes \cite{boyd2020nonlinear}) leads to a contribution to the signal field $\hat{E}^{(i)}$ which is linear in $\chi^{(2)}$:
\begin{multline} \label{eq:FirstOrderr}
\hat{E}_{s}^{(i)}(\vec{r}, t) = -2 \chi^{(2)}   \hspace{-.2cm}\int_{\vec{r}^\prime,t^\prime} \hspace{-.4cm} \mathcal{R}^{(i)}(\Greektens{$\rho$},\tau) \\
\times \mathcal{E}^{(i)}(\vec{r}^\prime, t^\prime)  \hat{ E}_\mathrm{vac}(\vec{r}^\prime, t^\prime).
\end{multline}
Here, \mbox{$\int_{\vec{r}^\prime,  t^\prime} = \int_{V_C}\dif^3 \vec{r}^\prime \int_{-\infty}^\infty \dif t^\prime$}, and $\mathcal{R}^{(i)}(\Greektens{$\rho$}, \tau)$ is the linear response function of the signal field, i.e., the classical Green tensor that propagates the field from a source at $\vec{r}^\prime, t^\prime$ to another space-time point $\vec{r}, t$, see also Eq.~\eqref{eq:ResponseFunction} and Appendix \ref{app:Prelim}. The generated NIR signal fields $\hat{E}_{s}^{(1)}$ and $\hat{E}_{s}^{(2)}$ are correlated only because there exist correlations in the quantum vacuum field $\hat{ E}_\mathrm{\mathrm{vac}}$. Inserting Eq.~\eqref{eq:FirstOrderr} for both $i=1, 2$ into Eq.~\eqref{eq:Signal} thus leads to the EOS signal stemming from vacuum field correlations:
\begin{align} \label{eq:EOSSignalVF}
G_{\theta_1\theta_2}\big|_{\mathrm{vac}}  =P_\mathrm{vac} \!\! \int_{\vec{r}, \vec{r}^\prime, t, t^\prime}\hspace{-0.8cm} \! L_1(\vec{r}, t)L_2(\vec{r}^\prime, t^\prime)\mathcal{C}_\mathrm{THz}(\Greektens{$\rho$}, \tau) .
\end{align}
Here, \mbox{$ P_\mathrm{vac}=  \mathrm{Im}[P(\theta_1)]\mathrm{Im}[P(\theta_2)]$}. Equation \eqref{eq:EOSSignalVF} has the same structure as the vacuum field contribution to $G^{(AB)}$ in Fermi's two atom setup, compare Eq.~\eqref{eq:SymmOrdVF}, and shows that $G_{\theta_1\theta_2}\big|_{\mathrm{vac}} $ is obtained by averaging the vacuum two-point correlation function of the vacuum field $\mathcal{C}$ over the spatio-temporal profiles of the two laser pulses $L_i$. For $\theta_1 = \theta_2 = \pi/2$ this result has also been obtained in Ref.~\cite{settembrini2022detection}.

\subsection{Source-Radiation Contribution}

In the second process which correlates $\hat{ E}^{(1)}$ and $\hat{ E}^{(2)}$, the broadband laser pulses generate source radiation $\hat{E}_\mathrm{s}$ via, e.g., difference frequency generation. In close analogy to Eq.~\eqref{eq:FirstOrderr}, $\hat{E}_\mathrm{s}$ is given by
\begin{multline}\label{eq:FirstOrderTHz}
\hat{E}_{s}(\vec{r},t) =-2\chi^{(2)} \hspace{-0.2cm} \int_{r^\prime, t^\prime} \hspace{-0.4cm}  \mathcal{R}(\Greektens{$\rho$}, \tau)  \mathcal{E}^{(i)}(\vec{r}^\prime,t^\prime)  \hat{E}_\mathrm{vac}^{(i)}(\vec{r}^\prime, t^\prime),
\end{multline}
where $\mathcal{R}$ is the response function of the field $\hat{E}$ defined in Eq.~\eqref{eq:ResponseFunction}. The source field $\hat{E}_{s}$ can propagate to the other laser pulse $\bar{i}$ where it can interact with it via the nonlinear coupling. This process thus relies on the exchange of source radiation between the two laser pulses and leads to the following contribution to the EOS signal (see Appendix \ref{app:EOS} for details)
\begin{multline}\label{eq:EOSSignalSRPrimes}
G_{\theta_1\theta_2} \big|_{s} = -\frac{\hbar}{2} \int_{\vec{r}, \vec{r}^\prime, t, t^\prime}\hspace{-0.8cm}       L_{1}(\vec{r}, t)L_{2}(\vec{r}^\prime, t^\prime) \\
\times [  P_\mathrm{SR}^\prime \mathcal{R}^\prime (\Greektens{$\rho$},\tau)+P_\mathrm{SR}^{\prime\prime} \mathcal{R}^{\prime\prime} (\Greektens{$\rho$},\tau)] ,
\end{multline}
with $P^\prime_\mathrm{s} = \mathrm{Im}[P(\theta_1)P(\theta_2)] $, $P^{\prime\prime}_\mathrm{s} = \mathrm{Im}[P(\theta_1) P^\ast(\theta_2)] $ and we introduced the symmetric (reactive) $ \mathcal{R}^{\prime} (\Greektens{$\rho$},\tau) = [ \mathcal{R}(\Greektens{$\rho$},\tau)  +  \mathcal{R}(\Greektens{$\rho$},-\tau) ]/2$ and antisymmetric (dissipative) $ \mathcal{R}^{\prime\prime} (\Greektens{$\rho$},\tau) = [ \mathcal{R} (\Greektens{$\rho$},\tau)  -  \mathcal{R} (\Greektens{$\rho$},-\tau) ]/2$ part of the response function.

\subsection{Individually Probing Source and Vacuum Fields in Space and Time}

We find that up to second order in $\chi^{(2)}$ the vacuum-field and source-radiation contributions in Eqs.~\eqref{eq:EOSSignalVF} and \eqref{eq:EOSSignalSRPrimes}, respectively, are the only ones such that $G_{\theta_1\theta_2}  = G_{\theta_1\theta_2} \big|_\mathrm{vac} + G_{\theta_1\theta_2}\big|_\mathrm{s}$. Tuning $\theta_1$ and $\theta_2$, all three contributions $G_{\frac{\pi}{2}\frac{\pi}{2}} = G_{\frac{\pi}{2}\frac{\pi}{2}} |_{\mathrm{vac}} \equiv G_\mathrm{vac}$ (only vacuum correlations [Eq.~\eqref{eq:EOSSignalVF} with $P_\mathrm{vac}=1$]), $G_{\frac{2\pi}{3}\frac{2\pi}{3}}-G_{\frac{4\pi}{3}\frac{4\pi}{3}} \equiv 2 G_{\mathcal{R}^\prime}$, and $G_{\frac{\pi}{2} \pi}-G_{\pi \frac{\pi}{2}} \equiv 2 G_{\mathcal{R}^{\prime\prime}}$ (only source radiation described by the symmetric and antisymmetric part of the response function [first and second term in Eq.~\eqref{eq:EOSSignalSRPrimes}], respectively) can be individually probed. Note that $G_{\mathcal{R}^{\prime}}$ and $G_{\mathcal{R}^{\prime\prime}}$ are obtained via the difference between two different measurements. For $\theta_1 =\pi/2$ and $\theta_2 = \pi$ one can also probe the full response function corresponding to source radiation propagating from mode $2$ to mode $1$, i.e., $G_{\frac{\pi}{2} \pi} \equiv  G_\mathrm{s}$ with
\begin{align}\label{eq:EOSSignalSR}
G_\mathrm{s}  = -  \frac{\hbar}{2}  \int_{\vec{r}, \vec{r}^\prime, t, t^\prime}\hspace{-0.8cm} \!\! L_{1}(\vec{r}, t)L_{2}(\vec{r}^\prime, t^\prime)   \mathcal{R}(\Greektens{$\rho$},\tau).
\end{align}  
The combinations of $\theta_1$ and $\theta_2$ which allow one to access the different contributions individually, are summarized in Fig.~\ref{fig:TwoLevelSystem}(c).  \\

\begin{figure}
\includegraphics[width=1.\columnwidth]{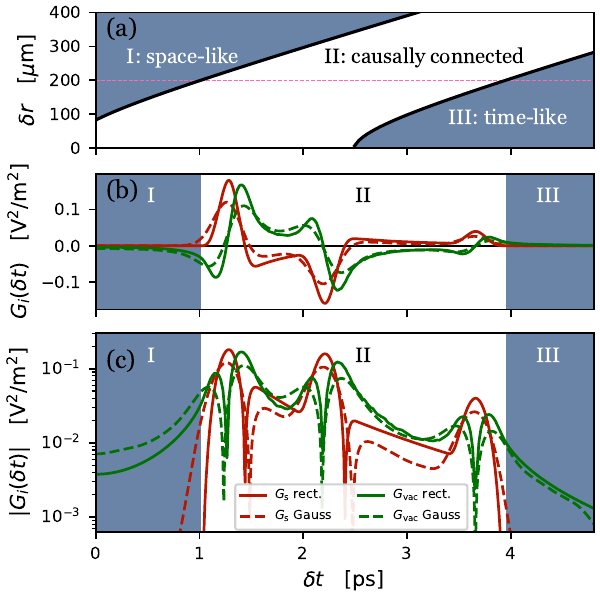}
\caption{\textit{Source radiation vs vacuum fluctuations contributions to the EOS signal}: (a) For THz refractive index $n = 3.33$, NIR group refractive index $n_g = 3.556$,  crystal length $L = 0.1\,$mm, and rectangular pulses with duration $\tau_\mathrm{p}= 185\, $fs and width $w= 10\,$\textmu m, we identify numerically (shaded areas) and analytically (black lines) the three different regions I (completely space-like), II (causal communication possible), and III (completely time-like) depending on the spatial distance $\delta r$ and delay $\delta t$ between the center of the pulses. (b) EOS signal from vacuum field fluctuations $G_{\mathrm{vac}}$ (red) or source radiation $G_{\mathrm{s}}$ (green) with rectangular (solid) or Gaussian shaped (dashed) pulses for  $\delta r  = 200\,$\textmu m [pink dashed line in (a)]. For the rectangular pulses we used the same parameters as in (a), for the Gaussian pulses we included dispersion and absorption effects, see main text. (c) Absolute values of the signals in (b) on a logarithmic scale.  }
\label{fig:EOSSignalDelt}
\end{figure}

Equations~\eqref{eq:EOSSignalVF} and \eqref{eq:EOSSignalSR} show that using EOS one can individually probe source and vacuum-field contributions, which are given by the correlation and response functions $\mathcal{C}$ and $\mathcal{R}$ averaged over space-time regions defined by the envelopes of the two laser pulses $L_{1,2}$, respectively. We next use this result to show how EOS experiments can reveal causality at the single photon level and space- and time-like correlations in the quantum vacuum field. The former implies that source radiation (and with it information transfer via a single photon) can only propagate at the finite speed of light inside the medium $c_n$. This is ensured by the fact that $\mathcal{R}(\Greektens{$\rho$}, \tau)$ is strictly zero if $\rho > c_n\tau$, i.e., only connects causally connected space-time points. The existence of correlations in the quantum vacuum between space- and time-like separated space-time regions implies $\mathcal{C}(\Greektens{$\rho$}, \tau) \neq 0$ even if $\rho > c_n\tau$ or $\rho < c_n\tau$, which leads to an EOS signal even if the two laser pulses are completely space- or time-like separated.

We first consider the idealized scenario in which the laser pulses have a rectangular shape and we can neglect dispersion and absorption inside the nonlinear crystal by assuming that the refractive index of the crystal in the THz $n(\Omega)$ is constant and real. We use $n(\Omega) = n =3.33$ similar to the one in GaP \cite{leitenstorfer1999detectors}, see Appendix \ref{app:Param} for details. In this case we find that the correlation and response functions are still given by Eqs.~\eqref{eq:CorrelationFree} and \eqref{eq:ResponseFree} subject to the replacement $c \to c_n $. Clearly if $\rho \gtrless c_n\tau$ we find $\mathcal{R}(\Greektens{$\rho$}, \tau) = 0$ whereas $\mathcal{C}(\Greektens{$\rho$}, \tau) \neq 0$. We assume that the rectangular laser pulses have a beam waist $ w = 10\,$\textmu m and duration $\tau_\mathrm{p} = 185\,$fs, as in the experiment in Ref.~\cite{settembrini2022detection}, and we set the length of the crystal to $L=0.1\,$mm [see Fig.~\ref{fig:TwoLevelSystem}(b)]. Using these parameters we can identify three different regimes [see Fig.~\ref{fig:EOSSignalDelt}(a)]: During the time the laser pulses are inside the crystal, they can (I) remain completely space-like separated (for large $\delta r > \delta r_{I/II}$), (II) be causally connected, i.e., a signal can propagate in the crystal with the speed of light from one to the other ($\delta r_{II/III}>\delta r > \delta r_{I/II}$), (III) remain completely time-like separated (for large delays $\delta t$ or $\delta r<\delta r_{II/III}$). We identify these three regions numerically in Fig.~\ref{fig:EOSSignalDelt}(a) (color code) and find that they are all within reach of current experimental setups \cite{settembrini2022detection}. The boundaries between the three regions can also be found analytically: if $\delta r \gg w$ we have $ \delta r_{I/II} =w + \sqrt{(c_n[\delta t + \tau_\mathrm{p}] + L n_g/n)^2  -  L^2} $, and for $c_n \delta t \gg w$ and $\delta r > w$ we find $\delta r_{II/III} = -w + \sqrt{(c_n[\delta t -\tau_\mathrm{p}] - L^2 n_g/n)^2-L^2} $, see black solid lines in Fig.~\ref{fig:EOSSignalDelt}(a). 

In Fig.~\ref{fig:EOSSignalDelt}(b,c) we show the vacuum-field and source-radiation contribution to the EOS signal $G_\mathrm{vac}$ and $G_\mathrm{s}$, obtained via Eqs.~\eqref{eq:EOSSignalVF} and \eqref{eq:EOSSignalSR}, respectively, as a function of $\delta t$ for $\delta r = 200\,$\textmu m, see Appendix \ref{app:EOSRec} for details. As expected, in regions (I) and (III) the source-radiation contribution is \emph{exactly} zero, i.e., $G_\mathrm{s} = 0$, whereas in region (II) we find $G_\mathrm{s} \neq 0$ indicating that the two laser pulses can exchange source radiation. The contribution from vacuum fluctuations $G_\mathrm{vac}$, however, is non-zero in all three regions (I)-(III). This shows that the two laser pulses can become correlated although they remain completely space- or time-like separated and thus cannot exchange source radiation.

In a next step, we turn to the more realistic scenario of two Gaussian laser pulses with the same beam waist $w$ and duration $\tau_\mathrm{p}$ as used for the rectangular pulses above. Also, we account for dispersion and (linear) absorption effects, by considering the complex valued, dispersive permittivity of GaP in the THz frequency range $\epsilon(\omega)$ as measured in Ref.~\cite{leitenstorfer1999detectors}. Using macroscopic QED \cite{scheel2009macroscopic} we find the response and correlation function accounting for the full polaritonic quantum vacuum in the presence of absorption and dispersion \cite{lindel2020theory,lindel2021macroscopic}, see Appendix \ref{app:Prelim} and \ref{app:Gauss}. As can be seen in Fig.~\ref{fig:EOSSignalDelt}(b,c), we find a very similar result for the EOS signal compared to the dispersion- and absorptionless case with two rectangular pulses considered above. The main features are slightly washed out by the frequency dependence of the speed of light due to the dispersive refractive index. Furthermore, the exponentially decaying tails of the Gaussian pulses cause the source radiation contribution to also decrease exponentially in regions I and III which is in sharp contrast to the algebraic decay of the vacuum fluctuation contribution. Thus, also for Gaussian pulses and including dispersion and absorption effects, one can identify the regions (I) and (III) in which the two pulses become correlated due to harvesting of correlations from the quantum vacuum, although their ability to exchange source radiation is exponentially suppressed.   \\

\begin{figure}
\includegraphics[width=1.\columnwidth]{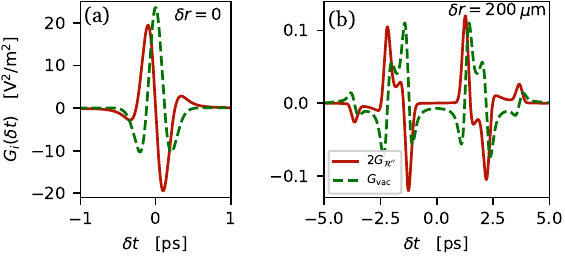}
\caption{\textit{Time-domain fluctuation--dissipation theorem}: For the same parameters as the dashed lines in Fig.~\ref{fig:EOSSignalDelt}(b,c), the EOS signals stemming from vacuum fluctuations $G_\mathrm{vac} $ (green dashed) and from the dissipative part of the response function $2G_{\mathcal{R}^{\prime\prime}}$ (solid red) are shown for spatial separations of the pulses of (a) $\delta r = 0$ and (b)  $\delta r =200\,$\textmu m. According to the time-domain FDT $2G_{\mathcal{R}^{\prime\prime}}(\delta t)$ is the Hilbert transform of $G_\mathrm{vac}(\delta t) $, compare Eq.~\eqref{eq:EOSTFDT}.  }
\label{fig:TFDT}
\end{figure}

\subsection{Fluctuation--Dissipation Theorem} \label{sec:FDT} 

The fluctuation--dissipation theorem (FDT) in frequency domain connects the dissipative part of the response function to the correlation function and reads (at zero temperature) 
\begin{align} \label{eq:FDTFreq}
\mi \mathcal{C}(\Greektens{$\rho$}, \omega) = \hbar \mathrm{sgn}(\omega) \mathcal{R}^{\prime\prime}(\Greektens{$\rho$}, \omega),
\end{align}
where $\mathrm{sgn}[x]$ is the sign function. In time domain it reads \cite{pottier_quantum_2001}
\begin{align} \label{eq:TFDT}
 \mathcal{C}(\tau) = -\frac{\hbar}{\pi} \mathcal{P} \int_{-\infty}^\infty \dif \tau^\prime \frac{\mathcal{R}^{\prime\prime}(\tau^\prime)}{\tau- \tau^\prime}  = - \hbar \mathcal{H} \{\mathcal{R}^{\prime\prime}(\tau)   \}.
\end{align} 
Here, $\mathcal{H}$ denotes the Hilbert transform. For the EOS signal this implies (see Appendix \ref{sec:FDTApp})
\begin{align} \label{eq:EOSTFDT}
G_{\mathrm{vac}}(\delta t) = \frac{2}{\pi} \mathcal{P} \int_{-\infty}^\infty \dif \delta t^\prime \frac{G_{\mathcal{R}^{\prime \prime}}(\delta t^\prime)}{\delta t- \delta t^\prime} .
\end{align}
We thus find a direct and somewhat surprising implication of the time-domain FDT for two-beam EOS: The correlations harvested from vacuum field fluctuations for a given time delay $\delta t$ can be obtained via a Hilbert transformation from the distribution of correlations for different $\delta t$ due to the antisymmetric part of the response function, and vice versa. This also means that EOS can be used to probe the FDT inherently in time-domain. Equation \eqref{eq:EOSTFDT} is illustrated in Fig.~\ref{fig:TFDT}, where we display $G_{\mathcal{R}^{\prime \prime}}(\delta t)$ and $G_{\mathrm{vac}}(\delta t)$ for two different values of $\delta r$. \\
Being able to probe the FDT locally in time, i.e., as a function of $\delta t$, implies that it is probed over a broad range of frequencies at once. As discussed in previous works \cite{benea-chelmus_electric_2019,lindel2020theory,lindel2021macroscopic}, Fourier transforming the EOS signal with respect to $\delta t$ gives access to the correlation and response function in frequency domain. Defining $G_i(\Omega) \equiv \frac{1}{2\pi} \int_{-\infty}^\infty \dif \delta t \, \me^{\mi \Omega \delta t }G_i(\delta t)  $, with $i = \mathrm{vac}, \mathrm{s}, \mathcal{R}^{\prime \prime}$, the frequency-domain fluctuation--dissipation theorem in Eq.~\eqref{eq:FDTFreq} implies
\begin{align}
\mathrm{sgn}(\Omega)  G_{\mathrm{vac}}(\Omega) & = 2 \mi \, G_{\mathcal{R}^{\prime\prime}}(\Omega) = -2  \mathrm{Im} G_{\mathrm{s}}(\Omega) .
\end{align}
By obtaining $G_{\mathcal{R}^{\prime\prime}}(\Omega)$ or $G_{\mathrm{s}}(\Omega)$ and $G_\mathrm{vac}(\Omega)$ from the experimental data for $G_{\mathcal{R}^{\prime\prime}}(\delta t)$ or $G_{\mathrm{s}}(\delta t)$ and $G_\mathrm{vac}(\delta t)$, one can thus also use EOS experiments for a broadband test of the frequency-domain fluctuation--dissipation theorem. The resolved spectral range in the THz is mainly determined by the temporal extent $\tau_\mathrm{p}$ of the lasers. For laser pulses with $\tau_\mathrm{p} = 185\,$fs as in Ref.~\cite{settembrini2022detection} this range is roughly $\Omega \in [0, 4]\,$THz.

\section{Conclusion and Outlook}

We have identified the contributions of the source and vacuum field in generating correlations in Fermi's two-atom setup. Our analysis revealed that these quantities can be uniquely identified by the constraint that they are individually hermitian and that the source-radiation contribution vanishes for space-like separated atoms. We further determined source and vacuum-field contributions in the EOS analog of Fermi's two-atoms setup. We have discussed the potential to use state-of-the-art EOS experiments to probe the causal nature of source radiation at the single photon level, as originally proposed by Fermi in 1932 \cite{fermi_quantum_1932}, the existence of space- and time-like correlations in the quantum vacuum, and the time-domain FDT. By expressing the EOS signal in terms of general correlation and response functions of the THz field, our findings can be extended to investigate the influence of complex environments such as cavities \cite{lindel2022probing}, thermal fluctuations, or material resonances. This approach also enables the analysis of how the space-time structure of correlations, present in THz quantum fields other than the quantum vacuum, can be probed in EOS experiments. In the future, it may be possible to create an analog of Fermi's two-atom setup in curved space-times by introducing an additional strong laser pulse to the nonlinear crystal, along with the two probe beams \cite{kizmann2019subcycle,drori2019observation,philbin_fiber-optical_2008}.\\

\hyphenation{Studien-stiftung}

\acknowledgements{We thank Heinz-Peter Breuer for fruitful discussions. F.L. acknowledges support from the Studienstiftung des deutschen Volkes. A.H. acknowledges financial support from Swiss National Science Foundation (SNSF) (grant 200020\_207795/1).}

\appendix

\section{Correlation and Response Functions of the Polaritonic Quantum Vacuum} 
\label{app:Prelim}

We review some basic definitions from macroscopic quantum electrodynamics and linear response theory.  \\

\subsection{Electric Field Operator and the Green Tensor} 

Using the framework of macroscopic quantum electrodynamics \cite{buhmann2013dispersion,scheel2009macroscopic}, the electric field operator in general dispersive and absorptive environments, described by a complex permittivity $\epsilon(\omega)$, is given by 
\begin{subequations} \label{eq:ElectricFieldOp}
\begin{align}
\hat{\vec{E}}_\mathrm{vac}(\vec{r}, t) & = \int_{0}^\infty \dif \omega \me^{-\mi \omega t} \hat{\vec{E}}_\mathrm{vac}(\vec{r}, \omega ) + \mathrm{h.c.}  , \\ \nonumber
\hat{\vec{E}}_\mathrm{vac}(\vec{r}, \omega ) & =  \mi \frac{\omega^2}{c^2} \sqrt{\frac{\hbar}{\pi \epsilon_0} \mathrm{Im}[\epsilon(\omega)] } \\  \label{eq:EvacFrequ}
& \quad \quad \quad \int \dif^3 r^\prime \tens{D}(\vec{r}, \vec{r}^\prime, \omega) \cdot \hat{\vec{f}}(\vec{r}^\prime, \omega).
\end{align}
\end{subequations}
Here, $\mathrm{h.c.}$ denotes the hermitian conjugate and $\hat{\vec{f}}$ and $\hat{\vec{f}}^\dagger$ are polaritonic annihilation and creation operators, respectively, satisfying the commutation relations
\begin{align}
[\hat{\vec{f}}(\vec{r}, \omega), \hat{\vec{f}}(\vec{r}^\prime, \omega^\prime)] &=  [\hat{\vec{f}}^\dagger(\vec{r}, \omega), \hat{\vec{f}}^\dagger(\vec{r}^\prime, \omega^\prime)]  = \tens{ 0}, \\
[\hat{\vec{f}}(\vec{r}, \omega), \hat{\vec{f}}^\dagger(\vec{r}^\prime, \omega^\prime)] & = \Greektens{$\delta$}(\vec{r}-\vec{r}^\prime) \delta(\omega-\omega^\prime).
\end{align}
Also, we made use of the classical Green tensor $\tens{D}$ of the vector Helmholtz equation defined via
\begin{align} \label{eq:GreensTensorDef}
\left(  \nabla \times \nabla \times - \frac{\omega^2}{c^2}\epsilon(\omega) \right) \tens{D}(\vec{r}, \vec{r}^\prime, \omega) =    \Greektens{$\delta$}(\vec{r}-\vec{r}^\prime)      ,
\end{align}
and the boundary condition $\tens{D}(\vec{r}, \vec{r}^\prime, \omega) \to 0 $ for \mbox{$|\vec{r}- \vec{r}^\prime |  \to \infty $}. A useful relation connects the two-point correlation function of the frequency domain field operator to the imaginary part of the Green tensor \cite{buhmann2013dispersion,scheel2009macroscopic}
\begin{multline} \label{eq:EvacEvactoGreens}
 \langle \hat{\vec{E}}_{\mathrm{vac}} (\vec{r}^\prime,\Omega) \hat{\vec{E}}_{\mathrm{vac}}^\dagger (\vec{r}^{\prime\prime},\Omega^\prime) \rangle  \\
  =   \frac{\hbar \mu_0}{\pi}\Omega^2 \delta(\Omega- \Omega^\prime)\mathrm{Im}[\tens{D}(\vec{r}^\prime, \vec{r}^{\prime\prime}, \Omega)].
\end{multline}
In the following, we only require the $xx$ component of the Green tensor $\mathsf{D}_{xx} \equiv \mathsf{D} $. In a general dispersive and absorptive bulk medium, $\mathsf{D}$ can be expressed as \cite{buhmann2013dispersion,scheel2009macroscopic}
\begin{multline} \label{eq:GreensGeneral}
\mathsf{D}(\vec{r}, \vec{r}^\prime, \omega) 
 =  \frac{\mi}{8\pi^2 } \int \dif^2 k_\parallel  \frac{\me^{\mi \vec{k}_\parallel \cdot (\vec{r}_\parallel - \vec{r}_\parallel)}}{k_z} \\
 \times  \left( 1- \frac{k_x^2}{k^2} \right) \me^{\mi k_z|z-z^\prime|} .
\end{multline}
In the case that the field propagates only in the positive $z$ direction, one can use the paraxial approximation by assuming $k_\parallel \ll k$ \cite{calvo2006quantum}, which leads to
\begin{align} \label{eq:GreensParaxial} 
\mathsf{D}(\vec{r}, \vec{r}^\prime, \omega) =  \frac{\mi}{2 k} \delta(\vec{r}_\parallel -\vec{r}_\parallel^\prime) \me^{\mi k (z-z^\prime)} .
\end{align}

For the case of lossless and dispersionless media, i.e., media with a real-valued constant permittivity $\epsilon(\omega) = \epsilon \in \mathbb{R}$, one can find an expansion of the $x$-polarized electric field in normal modes \cite{blow1990continuum,loudon2000quantum}
\begin{multline} \label{eq:NMElectricField}
\hat{E}_{\mathrm{vac}} (\vec{r},t) = \mi \int \dif^3 k \sqrt{\frac{\hbar  \omega_k}{16 \pi^3 \epsilon_0  n^2}} \\
\times \left( \sum_{\lambda = 1,2} e_x(\vec{k}, \lambda) \hat{a}(\vec{k}, \lambda) \me^{-\mi \omega_k t + \mi \vec{k} \cdot \vec{r}} + \mathrm{h.c.} \right) ,
\end{multline}
where $\vec{e}(\vec{k}, 1)$ and $\vec{e}(\vec{k},2)$ are two transverse polarization vectors, $n = \sqrt{\epsilon}$ is the constant refractive index, and $\omega_k = c_n k \equiv c k/n$.

\subsection{Response Function} 

The response function $\mathcal{R}(\vec{r}, \vec{r}^\prime,t-t^\prime) $ of one polarization direction (given by a unit basis vector $\vec{e}_i$) of the electric field operator $\hat{\vec{E}} \cdot \vec{e}_i \equiv \hat{E}$ is given in Eq.~\eqref{eq:ResponseFunction} of the main text. The Fourier transform of this quantity is proportional to the classical Green tensor defined in Eq.~\eqref{eq:GreensTensorDef}:
\begin{align}  \label{eq:ResponseTimetoFreq}
\mathcal{R}(\vec{r}, \vec{r}^\prime, \omega) & = \frac{1}{2\pi} \int \dif \tau \, \me^{\mi \omega \tau }\mathcal{R}(\vec{r}, \vec{r}^\prime, \tau) \\
& =\frac{\mu_0 \omega^2}{2\pi} \mathsf{D}( \vec{r}, \vec{r}^\prime \omega),
\end{align}
or 
\begin{align}
\mathcal{R}(\vec{r}, \vec{r}^\prime, \tau) 
 & =   \int \dif \omega \, \me^{-\mi \omega \tau }  \frac{\mu_0 \omega^2}{2\pi} \mathsf{D}( \vec{r}, \vec{r}^\prime \omega).
\end{align}
These relations can be derived by inserting Eq.~\eqref{eq:ElectricFieldOp} into Eq.~\eqref{eq:ResponseFunction}. 
The response function describes the response of the field at time $t$ and position $\vec{r}$ to a source at time $t^\prime$ and position $\vec{r}^\prime$. One thus finds $\mathcal{R}(\vec{r}, \vec{r}^\prime,t-t^\prime) = 0 $ for $\tau \equiv t-t^\prime   < |\vec{r}-\vec{r}^\prime|/c$ as required by special relativity. As $\mathcal{R}$ is not symmetric under $\tau \to -\tau$, one can divide it into symmetric (reactive) $\mathcal{R}^\prime$ and antisymmetric (dissipative) $\mathcal{R}^{\prime\prime}$ parts:
\begin{align}
\mathcal{R}^{\prime\prime}(\tau) & = \frac{1}{2}[\mathcal{R}(\tau) - \mathcal{R}(-\tau)] \\ \label{eq:RDissip}
& =  \frac{\mi}{2\hbar}  [\hat{E}(\vec{r}, t), \hat{E}(\vec{r}^\prime, t^\prime)],\\
\mathcal{R}^{\prime}(\tau) & = \frac{1}{2}[\mathcal{R}(\tau) + \mathcal{R}(-\tau)].
\end{align}
In frequency domain one finds $\mathcal{R}^{\prime\prime}(\omega) = \mi \mathrm{Im}[\mathcal{R}(\omega)]$ and $\mathcal{R}^\prime(\omega) = \mathrm{Re}[\mathcal{R}(\omega)]$.

\subsection{Correlation Function}

The correlation function of one polarization direction of the electric field operator is defined in Eq.~\eqref{eq:CorrelationFunction}. In frequency space it can be evaluated via Eq.~\eqref{eq:EvacEvactoGreens} and one obtains
\begin{align} \label{eq:CorrelationFunctionFrequency}
\mathcal{C}(\vec{r}^\prime, \vec{r}^{\prime\prime}, \omega) =   \frac{\hbar \mu_0}{2\pi} \mathrm{sgn}[\omega]\omega^2\mathrm{Im}[\mathsf{D}(\vec{r}^\prime, \vec{r}^{\prime\prime}, \omega)],
\end{align}
where $\mathrm{sgn}[x] \equiv 2 \theta(x)-1$ is the sign function. The correlation function has the following symmetry properties: $\mathcal{C}(\tau) = \mathcal{C}(-\tau) $, $\mathcal{C}(\omega) = \mathcal{C}^\ast(-\omega) $. 

From Eq.~\eqref{eq:CorrelationFunction} and \eqref{eq:RDissip} one finds the useful relation
 \begin{align} \label{eq:Evac2inLinearResponse}
 \braket{\hat{E}_\mathrm{vac} (\vec{r}, t) \hat{E}_\mathrm{vac} (\vec{r}^\prime, t^\prime)} = \mathcal{C}(\Greektens{$\rho$}, \tau) - \mi \hbar \mathcal{R}^{\prime\prime}(\Greektens{$\rho$}, \tau).
 \end{align}

\subsection{Normal Mode QED}

In an absorptionless and dispersionless medium, i.e., in case the refractive index is constant and real, we can use the normal-mode expression for the quantized electric field in Eq.~\eqref{eq:NMElectricField}. This can be used to find the response and correlation functions via Eqs.~\eqref{eq:ResponseFunction} and \eqref{eq:CorrelationFunction}:
\begin{align} \label{eq:ResponseNormalMode}
\mathcal{R}(\Greektens{$\rho$}, \tau) & = \frac{ \mu_0}{4 \pi   } \square_n \frac{1}{\rho}  \delta\left(\frac{\rho}{c_n}-\tau \right) , \\\label{eq:CorrelationNormalMode}
\mathcal{C}(\Greektens{$\rho$}, \tau) & = \frac{\mu_0 \hbar}{8 \pi^2   } \square_n \frac{1}{\rho} \left(\frac{\mathcal{P}}{\frac{\rho}{c_n}- \tau} + \frac{\mathcal{P}}{\frac{\rho}{c_n}+ \tau}\right),
\end{align}
where we have also defined $\square_n \equiv \frac{\partial^2}{\partial t \partial t^\prime} - c_n^2 \frac{\partial^2}{\partial x \partial x^\prime}$. In case $n = 1$ they reduce to the free-space response and correlation functions of the electric field operator, which can be found in standard quantum optics textbooks, see, e.g., Ref.~\cite{cohen1997photons}, and Eqs.~\eqref{eq:ResponseFree} and \eqref{eq:CorrelationFree} of the main text. The space-time structure of $\mathcal{R}$ and $\mathcal{C}$ in Eqs.~\eqref{eq:ResponseNormalMode} and \eqref{eq:CorrelationNormalMode} is apparent: For $\tau < \rho/c_n$, $\mathcal{R}$ is strictly zero whereas $\mathcal{C}$ is not. Also, the response and correlation functions in Eqs.~\eqref{eq:ResponseNormalMode} and \eqref{eq:CorrelationNormalMode} satisfy the time-domain fluctuation dissipation theorem in Eq.~\eqref{eq:TFDT} as can be verified by using that $\mathcal{H}[\delta(\tau)]= \mathcal{P}/(\pi \tau)$.

\section{Details on Fermi's Two Atom Problem} \label{app:Fermi}

In Appendix \ref{app:FermiSingle}, we consider the equations of motion of single atom observables in Fermi's two-atom setup and, following Refs.~\cite{dalibard_vacuum_1982,dalibard_dynamics_1984}, discuss how contributions from the source and vacuum-field can be identified. In Appendix \ref{app:FermiCorr} we determine source radiation and vacuum-field contributions to the generation of correlations between the two atoms as discussed in Sec.~\ref{sec:FermiTwoAtomCorr} of the main text. This includes a derivation of Eqs.~\eqref{eq:SymmOrdVF} and \eqref{eq:SymmOrdSR}.

\subsection{Single-Atom Observables} \label{app:FermiSingle}

We consider the Hamiltonian of the Fermi two-atom setup, which reads
\begin{align}
\hat{H} =\hat{H}_A   +  \hat{H}_F+  \hat{H}_I^{(F)},
\end{align}
with $\hat{H}_F$ is the free field Hamiltonian, and $\hat{H}_A $ the free atomic Hamiltonian, which reads 
\begin{align}
\hat{H}_A = \frac{\hbar}{2} \omega_0 \sum_{i=A,B}  \hat{ \sigma}_z^{(i)} .
\end{align}
Furthermore, $\omega_0$ is the transition frequency of the two atoms, and $\hat{\sigma}_z^{(i)} = \hat{\sigma}_+^{(i)} \hat{\sigma}_-^{(i)}  - \hat{\sigma}_-^{(i)} \hat{\sigma}_+^{(i)}  $ with the raising and lowering operator $\hat{\sigma}_+^{(i)}$  and $\hat{\sigma}_-^{(i)} $ of atom $i$, respectively. $H_I^{(F)}$ is given in Eq.~\eqref{eq:IAHamiltonianFermi} with the dipole moment for a two-level system $\hat{d}^{(i)}  = d \hat{\sigma}_x^{(i)}$ with $\hat{\sigma}_x^{(i)} =\hat{\sigma}_-^{(i)} + \hat{\sigma}_+^{(i)}$. Before considering the equation of motion of a generic observable of atom $A$, we first solve the Heisenberg equations of motion of $\hat{E}$ and $\hat{d}^{(i)}$ up to first order in the interaction Hamiltonian and find
\begin{align} \label{eq:GeneralSourceFree1}
\hat{E}(\vec{r}, t) &\approx \hat{E}_\text{vac}(\vec{r}, t)  + \hat{E}_\text{s,A}(\vec{r}, t) +\hat{E}_\text{s,B}  (\vec{r}, t) , \\ \label{eq:GeneralSourceFree2}
\hat{d}^{(i)}(t) & \approx \hat{d}_\text{vac}^{(i)}(t) + \hat{d}_\text{s}^{(i)}(t) .
\end{align}
The solution in lowest order in $\hat{H}_I^{(F)}$, denoted by the subscript $\mathrm{vac}$, are just the solutions of the uncoupled field and dipole operator. The first order corrections are the source terms given by
\begin{multline} \label{eq:SourceAtomi}
 \hat{E}_\text{s,i}  (\vec{r}, t)  =  \int \dif^3 r^\prime F(\vec{r}^\prime- \vec{r}_i)\int_{t_0}^\infty  \dif t^\prime  \eta(t^\prime) \\
 \times \mathcal{R}(\vec{r}, \vec{r}^\prime, t -t^\prime) \hat{d}^{(i)}(t^\prime), 
\end{multline}
 and
\begin{multline}
  \hat{d}_\text{s}^{(i)} (t)  =  \int \dif^3 r F(\vec{r}- \vec{r}_i)\int_{t_0}^\infty  \dif t^\prime  \eta(t^\prime) \\
  \times  \mathcal{R}_d^{(i)}(t -t^\prime) \hat{E}(\vec{r}, t^\prime).
\end{multline}
 Here, we defined the linear response function of the dipole operator
\begin{align} \label{eq:LinearResponsed}
\mathcal{R}^{(i)}_d(t-t^\prime) = \frac{\mi}{\hbar} \theta(t-t^\prime) [\hat{d}^{(i)}( t), \hat{d}^{(i)}(t^\prime)].
\end{align}

Next, we consider the Heisenberg equation of motion for a general observable of atom $A$, $\hat{O}^{(A)}$, which reads
\begin{multline} \label{eq:OAtimeEvo}
\frac{\partial}{\partial t}\hat{O}^{(A)}(t)  = \frac{i}{\hbar} [\hat{H}_A, \hat{O}^{(A)}  (t)]   -\frac{i}{\hbar}\eta(t) \int \dif^3 r   \\
\times  F(\vec{r} - \vec{r}_A)  \mathcal{O}\{ \hat{E}(\vec{r}, t), [\hat{d}^{(A)}(t), \hat{O}^{(A)}(t)] \}. 
\end{multline}
Note, that as $\hat{E}(\vec{r}, t) $ and $[\hat{d}^{(i)}(t), \hat{O}^{(A)}(t)]$ commute, one can continue the calculation with different orderings of these expressions in Eq.~\eqref{eq:OAtimeEvo}. To account for that, we introduced an operator ordering function $\mathcal{O}$. For symmetric operator ordering, for example, it is given by
\begin{multline} \label{eq:SymmetricOO}
 \mathcal{O}\{ \hat{E}(\vec{r}, t), \hat{O}^{(A)} (t)\}   \\
  = \frac{1}{2} \{ \hat{E}(\vec{r}, t) \hat{O}^{(A)} (t) 
   + \hat{O}^{(A)}(t) \hat{E}(\vec{r}, t)\} ,
\end{multline}
and for normal operator ordering by
\begin{multline} \label{eq:NormalOO}
 \mathcal{O}\{ \hat{E}(\vec{r}, t), \hat{O}^{(A)}(t) \}   \\
  =  \hat{E}^{(+)}(\vec{r}, t) \hat{O}^{(A)}  (t)
   + \hat{O}^{(A)}(t) \hat{E}^{(-)}(\vec{r}, t) .
\end{multline}
Here, $ \hat{E}^{(+)}$ and $ \hat{E}^{(-)}$ are the positive and negative frequency field components. 
To obtain the equation of motion of $\hat{O}^{(A)}$ up to second order in the interaction Hamiltonian, we insert Eqs.~\eqref{eq:GeneralSourceFree1} and \eqref{eq:GeneralSourceFree2} into Eq.~\eqref{eq:OAtimeEvo} and only keep terms up to second order to find
\begin{widetext}
\begin{multline} \label{eq:OAtimeEvoSecond}
\frac{\partial}{\partial t}\hat{O}^{(A)}(t)  \approx \frac{i}{\hbar} [\hat{H}_A, \hat{O}^{(A)}  (t)]   -\frac{i}{\hbar}\eta(t) \int \dif^3 r   
   F(\vec{r} - \vec{r}_A)  \mathcal{O}\{ \hat{E}_\mathrm{vac}(\vec{r}, t), [\hat{d}^{(A)}_\mathrm{vac}(t), \hat{O}_\mathrm{vac}^{(A)}(t)] \} \\
 + \frac{\partial}{\partial t}\hat{O}^{(A)}(t)  \big|_\mathrm{vac} + \frac{\partial}{\partial t}\hat{O}^{(A)}(t)  \big|_\mathrm{s} . 
\end{multline}
Here, we identified second-order terms proportional to the vacuum field $\hat{E}_\mathrm{vac}$ and source field $ \hat{E}_\mathrm{s}$, which read
\begin{align} \label{eq:SingleAtomObsVac}
\frac{\partial}{\partial t}\hat{O}^{(A)}(t)  \big|_\mathrm{vac}  & =  -\frac{i}{\hbar}\eta(t) \int \dif^3 r   
  F(\vec{r} - \vec{r}_A)  \mathcal{O}\{ \hat{E}_\mathrm{vac}(\vec{r}, t), [\hat{d}^{(A)}_\mathrm{s}(t), \hat{O}_\mathrm{vac}^{(A)}(t)] +   [\hat{d}^{(A)}_\mathrm{vac}(t), \hat{O}_\mathrm{s}^{(A)}(t)]  \}, \\ \label{eq:SingleAtomObss}
 \frac{\partial}{\partial t}\hat{O}^{(A)}(t)  \big|_\mathrm{s}  & =  -\frac{i}{\hbar}\eta(t) \int \dif^3 r   
  F(\vec{r} - \vec{r}_A)  \sum_{i=A,B}\mathcal{O}\{ \hat{E}_\mathrm{s,i}(\vec{r}, t), [\hat{d}^{(A)}_\mathrm{vac}(t), \hat{O}_\mathrm{vac}^{(A)}(t)]  \},
\end{align}
respectively. In Eqs.~\eqref{eq:SingleAtomObsVac} and \eqref{eq:SingleAtomObss} we also introduced the zeroth and first order contribution of the observable $\hat{O}^{(A)}(t)$, i.e., $\hat{O}_\mathrm{vac}^{(A)}(t)$ and $\hat{O}_\mathrm{s}^{(A)}(t)$, respectively. In case that $[\hat{H}_A , \hat{O}^{(A)}] = -\hbar \tilde{\omega}  \hat{O}^{(A)}$, with some frequency $\tilde{\omega}$, we find $  \hat{O}_\mathrm{vac}^{(A)}(t) = \me^{-\mi \tilde{\omega} (t-t_0)} \hat{O}_\mathrm{vac}^{(A)}(t_0)$ and 
\begin{align} \label{eq:Os}
  \hat{O}_\text{s}^{(A)} (t)  = -\frac{ \mi }{\hbar} \int \dif^3 r F(\vec{r}- \vec{r}_A)\int_{t_0}^t \dif t^\prime  \eta(t^\prime)  [\hat{d}^{(A)}_\mathrm{vac}(t^\prime), \hat{O}_\mathrm{vac}^{(A)}(t) ] \hat{E}_\mathrm{vac}(\vec{r}, t^\prime).
\end{align}

We integrate Eq.~\eqref{eq:OAtimeEvoSecond} assuming $[\hat{H}_A , \hat{O}^{(A)}] = -\hbar \tilde{\omega}  \hat{O}^{(A)}$ and we also take the vacuum expectation value with respect to the field degree of freedom $\braket{\cdot}_E$ and obtain
\begin{align} \label{eq:SolOA}
\braket{\hat{O}^{(A)}(t) }_E=\braket{ \hat{O}_\mathrm{vac}^{(A)}(t)  }_E + \braket{ \hat{O}^{(A)}(t)}_E\big|_\mathrm{s} +\braket{ \hat{O}^{(A)}(t)}_E\big|_\mathrm{vac},
\end{align}
with 
\begin{multline}  \label{eq:SingleAtomObsVacInt}
\braket{  \hat{O}^{(A)}(t)}_E\big|_\mathrm{vac}  =  -\frac{i}{\hbar} \int_{t_0}^{t} \dif t^\prime \me^{-\mi \tilde{\omega}(t-t_0)}  \eta(t^\prime) \int \dif^3 r   
   F(\vec{r} - \vec{r}_A)  \\
 \braket{\mathcal{O}\{ \hat{E}_\mathrm{vac}(\vec{r}, t^\prime), [\hat{d}^{(A)}_\mathrm{s}(t^\prime), \hat{O}_\mathrm{vac}^{(A)}(t^\prime)] +   [\hat{d}^{(A)}_\mathrm{vac}(t^\prime), \hat{O}_\mathrm{s}^{(A)}(t^\prime)]  \} }_E,  
\end{multline}
and  
 \begin{align}   \label{eq:SingleAtomObssInt}
   \braket{  \hat{O}^{(A)}(t)}_E\big|_\mathrm{s} &  =  -\frac{i}{\hbar} \int_{t_0}^{t} \dif t^\prime \me^{-\mi \tilde{\omega}(t-t_0)} \eta(t^\prime) \int \dif^3 r   
  F(\vec{r} - \vec{r}_A) \sum_{i=A,B} \braket{ \mathcal{O}\{ \hat{E}_\mathrm{s,i}(\vec{r}, t^\prime), [\hat{d}^{(A)}_\mathrm{vac}(t^\prime), \hat{O}_\mathrm{vac}^{(A)}(t^\prime)]  \} }_E .
\end{align}
\end{widetext}
Equations~\eqref{eq:SingleAtomObsVacInt} and \eqref{eq:SingleAtomObssInt} are the vacuum and source-field contribution to the atoms dynamics, respectively. One can analyze the above equations for different observables $\hat{O}^{(A)}$ and initial states of the atom to find the vacuum-field and source-radiation contribution to different physical processes. For example, spontaneous emission can be analyzed if the atom is initially excited and using $\hat{O}^{(A)} = \hat{\sigma}_z^{(A)}$ \cite{dalibard_vacuum_1982}. The only term in Eq.~\eqref{eq:SolOA}, which depends on the state of atom $B$, is the term proportional to $\hat{E}_\mathrm{s,B}$ in the source radiation contribution in Eq.~\eqref{eq:SingleAtomObssInt}. As also discussed in the main text, atom $A$ thus only notices the presence of atom $B$ through the source radiation emitted by atom $B$. As $\hat{E}_\mathrm{s,B}(\vec{r}, t^\prime)$ is independent of the state of atom $A$, it commutes with $ [\hat{d}^{(A)}_\mathrm{vac}(t^\prime), \hat{O}_\mathrm{vac}^{(A)}(t^\prime)]$, making this contribution independent of the chosen operator ordering. 

\subsection{Correlations} \label{app:FermiCorr}

In this section we obtain the source-radiation and vacuum-field contribution to the two-point correlation function $G^{(AB)}$. The calculation follows along similar lines as the one for single-atom observables in the last section. For simplicity, we assume that the two atoms are initially uncorrelated, i.e., $G^{(AB)}(t_0) = 0$.

We consider the Heisenberg equation of motions of the two-point correlation function $G^{(AB)}(t)$ defined in Eq.~\eqref{eq:FermiCorrelationFunction}. The first term on the right hand side of Eq.~\eqref{eq:FermiCorrelationFunction} evolves according to 
\begin{multline} \label{eq:OOFirst}
\frac{\partial}{\partial t} \hat{O}^{(AB)}(t)  =\frac{\mi}{\hbar} [\hat{H}_A, \hat{O}^{(AB) }(t)] 
 - \frac{\mi  }{\hbar} \eta(t) \int \dif^3 r  \\
  \times \sum_{i=A,B}  F(\vec{r}- \vec{r}_i) \mathcal{O}\left\{  \hat{E}(\vec{r}, t) , [\hat{d}^{(i)}(t), \hat{O}^{(AB)}(t) ]  \right\} .
\end{multline}
Here, we introduced the shorthand notation $\hat{O}^{(AB)}(t) = \hat{O}^{(A)}(t) \hat{O}^{(B)}(t)$. We insert Eqs.~\eqref{eq:GeneralSourceFree1} and \eqref{eq:GeneralSourceFree2} into Eq.~\eqref{eq:OOFirst} and only keep terms up to second order in the interaction Hamiltonian to find
\begin{multline} \label{eq:OOEOM2}
\frac{\partial}{\partial t} \hat{O}^{(AB)}(t)   =\frac{\mi}{\hbar} [\hat{H}_A, \hat{O}^{(AB) }(t) ]   - \frac{\mi  }{\hbar} \eta(t) \int \dif^3 r    \\
  \times  \sum_{i=A,B} F(\vec{r}- \vec{r}_i) \mathcal{O}\left\{  \hat{E}_\mathrm{vac}(\vec{r}, t) , [\hat{d}_\mathrm{vac}^{(i)}(t), \hat{O}_\mathrm{vac}^{(AB)}(t) ]  \right\} \\
 +  \frac{\partial}{\partial t} \hat{O}^{(AB)}(t) \big|_\mathrm{vac}+ \frac{\partial}{\partial t} \hat{O}^{(AB)}(t) \big|_\mathrm{s}.
\end{multline}
We have again identified vacuum-field and source-radiation contributions as before, which read
\begin{widetext}
\begin{align}
\frac{\partial}{\partial t} \hat{O}^{(AB)}(t) \big|_\mathrm{vac} & = - \frac{\mi  }{\hbar} \eta(t) \int \dif^3 r \sum_{i=A,B} F(\vec{r}- \vec{r}_i) \mathcal{O}\left\{  \hat{E}_\mathrm{vac}(\vec{r}, t) , [\hat{d}_\mathrm{s}^{(i)}(t), \hat{O}_\mathrm{vac}^{(AB)}(t) ] + [\hat{d}_\mathrm{vac}^{(i)}(t), \hat{O}_\mathrm{s}^{(AB)}(t) ]  \right\} , \\
\frac{\partial}{\partial t} \hat{O}^{(AB)}(t) \big|_\mathrm{s} & = - \frac{\mi  }{\hbar} \eta(t) \int \dif^3 r \sum_{i=A,B} F(\vec{r}- \vec{r}_i) \mathcal{O}\left\{  \hat{E}_\mathrm{s}(\vec{r}, t) , [\hat{d}_\mathrm{vac}^{(i)}(t), \hat{O}_\mathrm{vac}^{(AB)}(t) ]\right\} .
\end{align}
 Here, assuming that 
\begin{align} \label{eq:OOHAComm}
[\hat{H}_A , \hat{O}^{(A)}  \hat{O}^{(B)}] = -\hbar (\tilde{\omega}^{(A)}  + \tilde{\omega}^{(B)}  )\hat{O}^{(A)}\hat{O}^{(B)},
\end{align}
we find $\hat{O}_\mathrm{vac}^{(AB)}(t) = \hat{O}_\mathrm{vac}^{(A)}(t)\hat{O}_\mathrm{vac}^{(B)}(t)$, and $\hat{O}_\mathrm{s}^{(AB)}(t) = \hat{O}_\mathrm{s}^{(A)}(t) \hat{O}_\mathrm{vac}^{(B)}(t) + \hat{O}_\mathrm{vac}^{(A)}(t) \hat{O}_\mathrm{s}^{(B)}(t)$ with $\hat{O}_\mathrm{s}^{(A)}(t) $ given in Eq.~\eqref{eq:Os} and $\hat{O}_\mathrm{s}^{(B)}(t) $ can be obtained by replacing $A \leftrightarrow B$ in Eq.~\eqref{eq:Os}. Using Eq.~\eqref{eq:OOHAComm}, we integrate Eq.~\eqref{eq:OOEOM2} and average over the field degree of freedom to find
\begin{align} \label{eq:OOIntegrated}
\braket{\hat{O}^{(AB)}(t)}_E  =  \hat{O}_\mathrm{vac}^{(AB)}(t) + \braket{\hat{O}^{(AB)}(t) }_E\big|_\mathrm{vac}  + \braket{\hat{O}^{(AB)}(t) }_E\big|_\mathrm{s}  ,
\end{align}
with 
\begin{multline}
\hat{O}^{(AB)}(t) \big|_\mathrm{vac}  = - \frac{\mi }{\hbar} \int_{t_0}^t \dif t^\prime \me^{-\mi (\tilde{\omega}^{(A)}  + \tilde{\omega}^{(B)}) (t-t^\prime)}  \eta(t^\prime) \int \dif^3 r \sum_{i=A,B} F(\vec{r}- \vec{r}_i) \\
\times  \mathcal{O}\left\{  \hat{E}_\mathrm{vac}(\vec{r}, t^\prime) , [\hat{d}_\mathrm{s}^{(i)}(t^\prime), \hat{O}_\mathrm{vac}^{(AB)}(t^\prime) ] + [\hat{d}_\mathrm{vac}^{(i)}(t^\prime), \hat{O}_\mathrm{s}^{(AB)}(t^\prime) ]  \right\} ,
\end{multline}
and 
\begin{align}
\hat{O}^{(AB)}(t) \big|_\mathrm{s}  = - \frac{\mi }{\hbar}  \int_{t_0}^t \dif t^\prime \me^{-\mi (\tilde{\omega}^{(A)}  + \tilde{\omega}^{(B)}) (t-t^\prime)}  \eta(t^\prime) \int \dif^3 r \sum_{i,j=A,B} F(\vec{r}- \vec{r}_i) \mathcal{O}\left\{  \hat{E}_\mathrm{s,j}\vec{r}, t^\prime) , [\hat{d}_\mathrm{vac}^{(i)}(t^\prime), \hat{O}_\mathrm{vac}^{(AB)}(t^\prime) ]\right\} .
\end{align}
The second term on the right hand side of Eq.~\eqref{eq:FermiCorrelationFunction} reads $\braket{\hat{O}^{(A)}} \braket{\hat{O}^{(B)}}$ and can be obtained by using the solution for $\hat{O}^{(A)}$ found in Eq.~\eqref{eq:SolOA}. An equivalent expression for $\braket{\hat{O}^{(B)}}$ is obtained by exchanging $A \leftrightarrow B$ in Eq.~\eqref{eq:SolOA}. Using this together with the result in Eq.~\eqref{eq:OOIntegrated} and remembering that the two atoms are initially uncorrelated, which also implies $\braket{\hat{O}^{(AB)}_\mathrm{vac}(t)} - \braket{\hat{O}^{(A)}_\mathrm{vac}(t)}\braket{\hat{O}^{(B)}_\mathrm{vac}(t)} = 0$, we find for the two-point correlation function
\begin{align}
G^{(AB)}(t) =  G^{(AB)}(t) \big|_\mathrm{vac} + G^{(AB)}(t) \big|_\mathrm{s},
\end{align}
with the vacuum and source-field contribution
\begin{multline} \label{eq:GABGerneralvac}
G^{(AB)}(t) \big|_\mathrm{vac}  = - \frac{\mi  }{\hbar} \int_{t_0}^t \dif t^\prime \me^{-\mi (\tilde{\omega}^{(A)}  + \tilde{\omega}^{(B)}) (t-t^\prime)}  \eta(t^\prime) \int \dif^3 r \sum_{i=A,B} F(\vec{r}- \vec{r}_i) \\
\times \braket{    \mathcal{O}\left\{  \hat{E}_\mathrm{vac}(\vec{r}, t^\prime) ,  [\hat{d}_\mathrm{vac}^{(i)}(t^\prime), \hat{O}_\mathrm{vac}^{(i)}(t^\prime) ]  \hat{O}_\mathrm{s}^{(\overline{i})}(t^\prime)  \right\}   } , 
\end{multline}
and 
\begin{multline} \label{eq:GABGernerals}
G^{(AB)}(t) \big|_\mathrm{s}  = - \frac{\mi  }{\hbar} \int_{t_0}^t \dif t^\prime \eta(t^\prime) \int \dif^3 r \sum_{i=A,B} F(\vec{r}- \vec{r}_i)    [\hat{d}_\mathrm{vac}^{(i)}(t^\prime), \hat{O}_\mathrm{vac}^{(i)}(t) ] \\
\times \left( \braket{\mathcal{O}\left\{  \hat{E}_\mathrm{s,\overline{i}}(\vec{r}, t^\prime) ,  \hat{O}_\mathrm{vac}^{(\overline{i})}(t)  \right\}     }  - \braket{\hat{E}_\mathrm{s,\overline{i}}(\vec{r}, t^\prime) } \braket{  \hat{O}_\mathrm{vac}^{(\overline{i})}(t) }  \right).
\end{multline}
\end{widetext}
Here, we defined $\overline{A} = B$ and $\overline{B} = A$. Choosing different operator orderings in Eqs.~\eqref{eq:GABGerneralvac} and \eqref{eq:GABGernerals} leads to different \textit{relative} contributions to the two-point correlation function stemming from source radiation and vacuum field fluctuations.

\paragraph{Symmetric ordering}

Using the symmetric operator ordering defined in Eq.~\eqref{eq:SymmetricOO} in Eqs.~\eqref{eq:GABGerneralvac} and \eqref{eq:GABGernerals}, we obtain the vacuum-field and source-radiation contributions to the two-point correlation function given in Eqs.~\eqref{eq:SymmOrdVF} and \eqref{eq:SymmOrdSR}, respectively. Equations~\eqref{eq:SymmOrdVF} and \eqref{eq:SymmOrdSR} can also be written as 
\begin{align} \label{eq:SymmOrdVFApp}
G^{(AB)}(t) \big|_\mathrm{vac}   = \frac{1}{2} \braket{\{ \hat{O}^{(A)}_\mathrm{s} (t), \hat{O}^{(B)}_\mathrm{s} (t) \} }
,
\end{align}
and 
\begin{multline} \label{eq:SymmOrdSRApp}
G^{(AB)}(t) \big|_\mathrm{s}  =\frac{1}{2} \sum_{i=A,B} \left[  \braket{\{ \hat{O}^{(i)}_2 (t), \hat{O}^{(\overline{i})}_\mathrm{vac} (t) \} } \right. \\
\left. -   \braket{\hat{O}^{(i)}_2 (t)} \braket{ \hat{O}^{(\overline{i})}_\mathrm{vac} (t)  }   \right]
\end{multline} 
We see from Eq.~\eqref{eq:SymmOrdVFApp} that the vacuum contribution is given by the correlations between the source terms $\hat{O}^{(A)}_\mathrm{s} $ and $\hat{O}^{(B)}_\mathrm{s} $. $\hat{O}^{(A)}_\mathrm{s} $ and $\hat{O}^{(B)}_\mathrm{s} $ describe the individual interaction of atom $A$ and atom $B$ with the vacuum field, compare Eq.~\eqref{eq:Os}. The source radiation contribution in Eq.~\eqref{eq:SymmOrdSRApp} is given by the correlations between the free space atomic operator $\hat{O}^{(i)}_\mathrm{vac}$ and $\hat{O}^{(i)}_2 (t)$. $\hat{O}^{(i)}_2 (t)$ is the contribution to $\hat{O}^{(i)}$ stemming from the process in which atom $i$ interacts with the source radiation emitted by atom $\overline{i}$. It is obtained by replacing $\hat{E}_\mathrm{vac}$ in Eq.~\eqref{eq:Os} by $\hat{E}_\mathrm{s}^{(\overline{i})}$. This is in line with the interpretation that the vacuum-field contribution arises due to the individual interaction of the atoms with the vacuum field and thereby swapping correlations present in the vacuum to the atoms, whereas the source-radiation contribution arises in case one atom interacts with the source radiation emitted by the other atom.

\paragraph{Normal ordering:}

Using the normal operator ordering defined in Eq.~\eqref{eq:NormalOO} in Eqs.~\eqref{eq:GABGerneralvac} and \eqref{eq:GABGernerals}, we obtain the following vacuum-field and source-radiation contributions to the two-point correlation function:
\begin{align} \label{eq:NormalOrdVFApp}
G^{(AB)}(t) \big|_\mathrm{vac}   = 0,
\end{align}
and 
\begin{multline} \label{eq:NormalOrdSRApp}
G^{(AB)}(t) \big|_\mathrm{s}   = \frac{1}{2} \braket{\{ \hat{O}^{(A)}_\mathrm{s} (t), \hat{O}^{(B)}_\mathrm{s} (t) \} }  + \frac{1}{2} \sum_{i=A,B} \\
\times  \left[  \braket{\{ \hat{O}^{(i)}_2 (t), \hat{O}^{(\overline{i})}_\mathrm{vac} (t) \} } -   \braket{\hat{O}^{(i)}_2 (t)} \braket{ \hat{O}^{(\overline{i})}_\mathrm{vac} (t)  }   \right].
\end{multline}
As expected, the sum of the vacuum and source field contribution using normal operator ordering [Eqs.~\eqref{eq:NormalOrdVFApp} and \eqref{eq:NormalOrdSRApp}] is the same as in the case of symmetric operator ordering [Eqs.~\eqref{eq:SymmOrdVFApp} and \eqref{eq:SymmOrdSRApp}]. However, the individual vacuum and source field contributions in Eqs.~\eqref{eq:SymmOrdVFApp} and \eqref{eq:NormalOrdVFApp} and in Eqs.~\eqref{eq:SymmOrdSRApp} and \eqref{eq:NormalOrdSRApp}, respectively, differ. While using symmetric operator ordering one obtains that correlations between the atoms arise due to source \textit{and} vacuum-field contributions, they only arise due to source radiation if one uses a normal operator ordering. 

To obtain Eq.~\eqref{eq:NormalOrdSRApp} we used that using normal ordering we find
\begin{widetext}
\begin{multline} \label{eq:NormalHelp}
\braket{\mathcal{O}\left\{  \hat{E}_\mathrm{s,i}(\vec{r}, t^\prime) ,  \hat{O}_\mathrm{vac}^{(i)}(t)  \right\}     } =  \int \dif^3 r^\prime F(\vec{r}^\prime- \vec{r}_i)\int_0^{t^\prime}  \dif t^{\prime\prime } \eta(t^{\prime\prime }) \\
\times  \left( \mathcal{R}^{(+)}(\vec{r}, \vec{r}^\prime, t^\prime -t^{\prime\prime }) \braket{ \hat{d}^{(i)}(t^{\prime\prime }) \hat{O}_\mathrm{vac}^{(i)}(t)   } + \mathcal{R}^{(-)}(\vec{r}, \vec{r}^\prime, t^\prime -t^{\prime\prime }) \braket{\hat{O}_\mathrm{vac}^{(i)}(t)  \hat{d}^{(i)}(t^{\prime\prime })  }   \right),
\end{multline}
where we defined
\begin{align}
 \mathcal{R}^{(\pm)} (\vec{r}, \vec{r}^\prime, t-t^\prime )  = \frac{\mi}{\hbar} \theta(t-t^\prime) [\hat{E}^{(\pm)}_\mathrm{vac}(\vec{r}, t),  \hat{E}_\mathrm{vac}(\vec{r}^\prime, t^\prime)].
\end{align}
$ \mathcal{R}^{(\pm)} $ satisfy the following relations to the response and correlation function
\begin{align} \label{eq:HelpApp1}
\mathrm{Re} \left[ \mathcal{R}^{(\pm)} (\vec{r}, \vec{r}^\prime, t-t^\prime )  \right] &= \frac{1}{2} \mathcal{R} (\vec{r}, \vec{r}^\prime, t-t^\prime ) \\ \label{eq:HelpApp2}
\mathrm{Im} \left[ \mathcal{R}^{(+)} (\vec{r}, \vec{r}^\prime, t-t^\prime )  \right]& = -\mathrm{Im} \left[ \mathcal{R}^{(-)} (\vec{r}, \vec{r}^\prime, t-t^\prime )  \right] =\frac{1}{\hbar} \theta(t-t^\prime) \mathcal{C} (\vec{r}, \vec{r}^\prime, t-t^\prime ) .
\end{align}
To obtain the last equality, we made use of the fluctuation--dissipation theorem in Eq.~\eqref{eq:TFDT}. To obtain Eq.~\eqref{eq:NormalOrdSRApp}, we use Eq.~\eqref{eq:HelpApp1} and \eqref{eq:HelpApp2} in Eq.~\eqref{eq:NormalHelp} and insert the resulting expression into Eq.~\eqref{eq:GABGernerals}. 

\paragraph{Other operator orderings}

We consider more generic operator orderings and find that in all cases other than the symmetric one, either the source-radiation contribution does not vanish for space-like separated atoms or the vacuum and source-radiation contributions are not real. First, we consider operator orderings of the form
\begin{align} \label{eq:General1}
 \mathcal{O}\{ \hat{E}(\vec{r}, t), \hat{O}^{(A)} (t)\}   
  =  \{ \lambda  \hat{E}(\vec{r}, t) \hat{O}^{(A)} (t) 
   + (1- \lambda) \hat{O}^{(A)}(t) \hat{E}(\vec{r}, t)\} ,
\end{align}
with $\lambda \in [0,1]$. For $\lambda =1/2$ we recover the symmetric operator ordering. Using Eq.~\eqref{eq:General1} in Eq.~\eqref{eq:GABGerneralvac} we find using Eq.~\eqref{eq:Evac2inLinearResponse}:
\begin{multline} \label{eq:SymmOrdVF2}
G^{(AB)}(t)\big|_\mathrm{vac} = \int \dif^3 r^{\prime\prime} \int_{t_0}^t \dif t^{\prime\prime} \int \dif^3 r^\prime \int_{t_0}^t \dif t^\prime  
 L^{(A)}(\vec{r}^\prime, t^\prime, t) L^{(B)}(\vec{r}^{\prime\prime}, t^{\prime\prime}, t) \left\{ \mathcal{C}(\vec{r}^\prime, \vec{r}^{\prime\prime} ,t^\prime-t^{\prime\prime}) \right.\\
+ \left. \mi  \left[ \lambda  \mathcal{R}^{\prime\prime}(\vec{r}^\prime, \vec{r}^{\prime\prime} ,t^\prime-t^{\prime\prime})   - (1- \lambda)  \mathcal{R}^{\prime\prime}(\vec{r}^\prime, \vec{r}^{\prime\prime} , t^\prime-t^{\prime\prime}) \right] \right\}.
\end{multline}
The second row vanishes if $\lambda = 1/2$, i.e., in case of symmetric operator ordering. For all other values of $\lambda$, it does not vanish in general and leads to a purely complex contribution to $G^{(AB)}(t)\big|_\mathrm{vac} $.  

Another type of operator orderings is 
\begin{multline} \label{eq:General2}
 \mathcal{O}\{ \hat{E}(\vec{r}, t), \hat{O}^{(A)} (t)\}   
  =  \{ \lambda \left[  \hat{E} ^{(+)} (\vec{r}, t) \hat{O}^{(A)} (t) +\hat{O}^{(A)} (t)   \hat{E} ^{(-)} (\vec{r}, t)   \right] \\
   + (1- \lambda) \left[ \hat{E} ^{(-)} (\vec{r}, t) \hat{O}^{(A)} (t) +\hat{O}^{(A)} (t)   \hat{E} ^{(+)} (\vec{r}, t)   \right]\} ,
\end{multline}
Here, again $\lambda \in [0,1]$ and we we recover the symmetric (normal) operator ordering for $\lambda =1/2$ ($\lambda = 1$). In this case we find that 
\begin{align} \label{eq:SymmOrdSR2}
G^{(AB)}(t)\big|_\mathrm{s} = \lambda  \left[ G^{(AB)}(t)\big|_\mathrm{s}^\mathrm{Sym} +G^{(AB)}(t)\big|_\mathrm{vac}^\mathrm{Sym}  \right] -  (1- \lambda)  \left[ G^{(AB)}(t)\big|_\mathrm{s}^\mathrm{Sym}   -  G^{(AB)}(t)\big|_\mathrm{vac}^\mathrm{Sym}    \right] .
\end{align}
Here, $ G^{(AB)}(t)\big|_\mathrm{vac}^\mathrm{Sym} $ and $G^{(AB)}(t)\big|_\mathrm{s}^\mathrm{Sym}$ are the vacuum and source-field contribution in Eqs.~\eqref{eq:SymmOrdVF} and \eqref{eq:SymmOrdSR}, respectively, obtained using symmetric operator ordering. For space-like separated atoms, we found that $G^{(AB)}(t)\big|_\mathrm{vac}^\mathrm{Sym}  \neq 0$ and $G^{(AB)}(t)\big|_\mathrm{s}^\mathrm{Sym} = 0$. Thus, the source-radiation contribution in Eq.~\eqref{eq:SymmOrdSR2} only vanishes in general for space-like separated atoms if $\lambda =1/2$.

\end{widetext}

\section{Electro-Optic Sampling Signal}  \label{app:EOS}

In this Appendix we include a detailed derivation of the EOS signal by perturbatively solving the equations of motion for the electric field operator emerging from the crystal. We further identify the contributions from source radiation and from vacuum field fluctuations using a symmetric operator ordering.

\subsection{Preliminaries}

\paragraph{Electric Field Contributions.}

As discussed in the main text, there are three different contributions to the electric field, which are relevant for the EOS signal. First, there are the two $y$-polarised, near infrared (NIR) probe pulses $\Eclo$ and $\Eclt$ which will be treated classically \footnote{The full $y$ polarized field in the vacuum picture is given by $\hat{E}_y =\sum_{i=1,2} \Ecli + \hat{E}_\mathrm{i, vac}$, where the scalar functions $\Ecli$ are the coherent, classical field amplitude and $\hat{E}_\mathrm{i, vac}$ is the contribution from vacuum fluctuations. As the vacuum fields co-propagating with the two laser pulses are uncorrelated ($\braket{ \hat{E}_\mathrm{1, vac} \hat{E}_\mathrm{2, vac} } =0$), we neglect $\hat{E}_\mathrm{i, vac}$. $\hat{E}_\mathrm{i, vac}$ lead, however, to shot noise on each balanced detector, respectively.} 
and are given by
\begin{align} \label{eq:laserPulse}
\Ecli(\vec{r}, t)& = \int_{-\infty}^\infty \dif \omega \, g_{i}(\vec{r}_\parallel) \Ecli(\omega) \me^{\mi \omega \left(\frac{n(\omega) z}{c} -t \right)}.
\end{align}
Here, we have defined the transverse mode function of the two laser pulses $g_1$ and $g_2$, respectively, which we assume to be equal but shifted by the spatial distance in transverse direction between the laser pulses $\delta \vec{r}_\parallel$, i.e., $g_{1}(\vec{r}_\parallel)  = g_{2}(\vec{r}_\parallel- \delta \vec{r}_\parallel)$. Furthermore, $\Eclo(\omega) $ and $\Eclt(\omega) $ have the form $ \Ecli(\omega) = \overline{\mathcal{E}}(\omega)  \me^{-\mi \delta t_i \omega }  $, with the normalized, real laser spectrum $\overline{\mathcal{E}}(\omega) $. Note that $\overline{\mathcal{E}}(\omega)  = \overline{\mathcal{E}}(-\omega) $. Also, we assume that the spectra are symmetric around the central laser frequency, i.e., $\overline{\mathcal{E}}(\omega - \omega_c) = \overline{\mathcal{E}}(\omega+ \omega_c) $, and that the width of the spectrum is much smaller then the central frequency of the laser pulses $\omega_c$. We approximate the wave vector $k(\omega) = n(\omega) \omega/c$ of the laser pulse using a Tailor expansion around the central frequency of the laser pulse $\pm\omega_c$ by assuming that the real refractive index $ n(\omega)$ in the NIR is sufficiently flat in the spectral range of the laser pulse. Introducing the group refractive index $n_g = c \partial k_\omega / \partial \omega |_{\omega_c}$ and defining $n(\omega_c) \equiv n_\mathrm{c}$ we obtain
\begin{multline}
\Ecli(\vec{r}, t) =   2 \mathrm{cos}\left[\omega_c \left( n_\mathrm{c} \frac{z}{c} - t  \right) \right]  g_{i}(\vec{r}_\parallel)  \\
\times \underbrace{ \int_{-\infty}^\infty \dif \omega \, \overline{\mathcal{E}}(\omega-\omega_c) \me^{-\mi \omega \left(t + \delta t_i - \frac{n_g z}{c} \right)}}_{\mathcal{E}\left(t+ \delta t_i-\frac{n_g z}{c}\right)}.
\end{multline}
This can be used to find the time-domain expression for the laser pulse
\begin{align} \label{eq:laserPulseTimDomain}
\Ecli(\vec{r}, t)&  = \sqrt{ 2\pi L\, L_i(\vec{r}, t)} 2 \mathrm{cos}\left[\omega_c \left( n_\mathrm{c} \frac{z}{c} - t - \delta t_i \right) \right],\\
L_i & = \frac{1}{2\pi L } \mathcal{E}^2\left(t + \delta t_i-\frac{n_g z}{c}\right) g_i^2(\vec{r}_\parallel) 
\end{align}
Here, $L$ is the crystal length, and $L_i$ are the pulse envelopes normalized such that 
\begin{align}
\int_{V_C} \dif^3 r \int_{-\infty}^\infty \dif t   L_1(\vec{r}, t) = 1,
\end{align}
where $V_C$ is the crystal volume, i.e., $\int_{V_C} \dif^3 r = \int_{-\infty}^\infty \dif x\int_{-\infty}^\infty \dif y \int_{-L/2}^{L/2} \,\dif z$. 

The second and third relevant part of the electric field are the $x$-polarised, quantized electric fields $\hat{E}_x$ which we split into a component in the THz ($\hat{E} $) and NIR  ($\Ei$) frequency ranges, i.e., $\hat{E}_x = \hat{E} + \Ei$. Without the nonlinear coupling, these fields can be expressed by their vacuum expressions
\begin{align} \label{eq:ETHzfrequ}
\hat{E}_\mathrm{vac} (\vec{r}, t) & = \int_{-\Lambda}^\Lambda \dif \Omega \, \me^{-\mi \Omega t}\hat{E}_\mathrm{vac}(\vec{r}, \Omega) , \\  \label{eq:NIRFull} 
\Ei_\mathrm{vac} (\vec{r}, t) & =  \int_{|\omega|> \Lambda} \dif \omega \, \me^{-\mi \omega t}\hat{E}_\mathrm{vac}^{(i)}(\vec{r}, \omega) .
\end{align}
$\hat{E}_\mathrm{vac}(\vec{r}, \Omega) $ is given in Eq.~\eqref{eq:EvacFrequ}, and we have introduced the transition frequency $\Lambda$ which separates THz frequencies from NIR ones. The NIR field $\Ei$ is, as the pulses, given in the paraxial approximation \cite{calvo2006quantum} such that $\hat{E}_\mathrm{vac}^{(i)}(\vec{r}, \omega)$ is given by Eq.~\eqref{eq:EvacFrequ} with the Green tensor in the paraxial approximation, see Eq.~\eqref{eq:GreensParaxial}. This can be used to find 
\begin{multline} \label{eq:ENIR}
\hat{E}_\mathrm{vac}^{(i)}(\vec{r}, \omega) = - \frac{\omega}{2c n_\mathrm{c}} \sqrt{\frac{\hbar}{\pi \epsilon_0} \mathrm{Im}\epsilon(\omega)   } \\ 
\int \dif z^\prime \, \me^{\mi k (z-z^\prime)} \hat{f}^{(i)}_x(\vec{r}_\parallel, z^\prime, \omega) .
\end{multline}
Note that the bosonic creation and annihilation operators in the NIR and in the THz, $ \hat{f}^{(i)}_x(\vec{r}, \omega)$ and $ \hat{f}_x(\vec{r}, \Omega)$, respectively, commute. Furthermore, the field operators of the two modes are uncorrelated, i.e., $\hat{f}^{(1)}_x(\vec{r}, \omega), \hat{f}^{(1) \dagger}_x(\vec{r}, \omega) $ and $\hat{f}^{(2)}_x(\vec{r}, \omega), \hat{f}^{(2) \dagger}_x(\vec{r}, \omega)  $ commute. This is assured in the two-beam experimental setup by generating the two laser pulses via a beam splitter from a single pulse, see Appendix C of Ref.~\cite{guedes_back_2022}.

\paragraph{Interaction Hamiltonian.}

The nonlinear coupling inside the nonlinear crystal introduces an effective coupling between the three different field components, which is given by the following interaction Hamiltonian \cite{onoe2022realizing}
\begin{align} \label{eq:HomiltonianGen}
\hat{H}_{I}(t) & =  \chi^{(2)} \sum_{i = 1,2} \int_{V_C} \dif ^3 r  \hat{E}_y(\vec{r}, t) \hat{E}_{x} (\vec{r}, t)  \hat{E}_{x} (\vec{r}, t) \\ \label{eq:Hamiltonian}
  & = 2 \chi^{(2)} \sum_{i = 1,2} \int_{V_C} \dif ^3 r  \Ecli(\vec{r}, t) \hat{E} (\vec{r}, t)  \Ei(\vec{r}, t) .
\end{align}
Here, we used $\hat{E}_x = \hat{E} +\sum_i \Ei$, and the fact that $\hat{E} $ and $\Ei$ commute. Furthermore, we have neglected contributions which are proportional to $\hat{E}^2$ or $\hat{E}^{(i)2}$, since they will be rapidly oscillating and are thus assumed to average to zero \footnote{The exponents of the phase factors $\me^{\mi t (2\Omega + \omega )}$ and $\me^{\mi t (\omega + \omega^\prime + \tilde{\omega}) }$ never add to something close to zero.}. We also have neglected terms proportional to $\Eclo \Et$ or $\Eclt \Eo$. If the two pulses are well separated in space, these terms do not contribute as there is no spatial overlap between $\Eclt$ and $\Eo$ or $\Eclo$ and $\Et$, respectively. But also if there is spatial overlap between the two pulses, these terms can be neglected, as in the experimental setup the two pulses propagate into slightly different directions and pulse $i$ can only efficiently couple to NIR fields which are co-propagating with it due to phase-matching constraints.

\subsection{Heisenberg Equation of Motions for the Fields}

To eventually obtain the EOS signal, we solve Heisenbergs equations of motion for the quantized fields $\Ei$ and $\hat{E}$ up to second order in the interaction Hamitlonian in Eq.~\eqref{eq:Hamiltonian}. We use this approach instead of the Dyson series approach employed in Refs.~\cite{lindel2020theory,lindel2021macroscopic,lindel2022probing} to highlight the similarities to the calculation of source radiation and vacuum-field contributions to the dynamics of atoms discussed in Appendix~\ref{app:Fermi}.

Solving the Heisenberg equations of motion for $\Ei$ and $\hat{E}$ up to first-order perturbation theory, we find $\hat{E} \approx \hat{E}_\mathrm{vac} + \hat{E}_s$ with the free fields defined in Eqs.~\eqref{eq:ETHzfrequ} and \eqref{eq:NIRFull}, and the source fields 
\begin{widetext}
\begin{align} \label{eq:EOMTimeNIR}
\Ei_s (\vec{r}, t) &  = -2 \chi^{(2)}  \int_{V_C} \dif ^3 r^\prime \int_{-\infty}^{\infty} \dif t^\prime \mathcal{R}^{(i)}( \vec{r}, \vec{r}^\prime,t-t^\prime) \, \mathcal{E}^{(i)}(\vec{r}^\prime, t^\prime) \hat{E}_\mathrm{vac}  (\vec{r}^\prime, t^\prime) , \\  \label{eq:EOMTimeTHz}
\hat{E}_s(\vec{r}, t) & =-2 \chi^{(2)} \sum_{i = 1,2} \int_{V_C} \dif ^3 r^\prime \int_{-\infty}^{\infty} \dif t^\prime \mathcal{R}( \vec{r}, \vec{r}^\prime,t-t^\prime) \, \Ecli\vec{r}^\prime, t^\prime) \Ei_\mathrm{vac}  (\vec{r}^\prime, t^\prime)  .
\end{align}
We used that $\Eo$ and $\Et$ commute and defined the response functions of the fields $\Ei$ and $\hat{E}$, namely $\mathcal{R}^{(i)}$ and $\mathcal{R}$, according to Eq.~\eqref{eq:ResponseFunction}. Transforming Eqs.~\eqref{eq:EOMTimeNIR} and \eqref{eq:EOMTimeTHz} to frequency space and using Eq.~\eqref{eq:ResponseTimetoFreq} we find
\begin{align}\label{eq:LikeDysonNIR}
\hat{E}_s^{(i)}(\vec{r}, \omega) & =- 2 \chi^{(2)} \mu_0 \omega^2   \int_{V_C} \dif ^3 r^\prime  \mathsf{D}(\vec{r}, \vec{r}^\prime, \omega) \int_{|\Omega| <\Lambda} \dif \Omega\,  \Ecli (\vec{r}^\prime, \omega-\Omega) \hat{E}_\mathrm{vac}  (\vec{r}^\prime,\Omega) , \\ \label{eq:LikeDysonTHz}
\hat{E}_s(\vec{r}, \Omega) & =- 2 \chi^{(2)} \mu_0 \Omega^2  \sum_{i = 1,2} \int_{V_C} \dif ^3 r^\prime   \mathsf{D}(\vec{r}, \vec{r}^\prime, \Omega) \int_{|\omega| >\Lambda} \dif \omega \, \Ecli(\vec{r}^\prime, \Omega-\omega) \Ei_\mathrm{vac}  (\vec{r}^\prime,\omega) .
\end{align}
We will also need the second order expression for the field $\Ei$ which is given by
\begin{align} \label{eq:EOMTimeNIR2}
\Ei_2 (\vec{r}, t) & = -2 \chi^{(2)}  \int_{V_C} \dif ^3 r^\prime \int_{-\infty}^{\infty} \dif t^\prime \mathcal{R}^{(i)}( \vec{r}, \vec{r}^\prime,t-t^\prime) \, \mathcal{E}^{(i)}(\vec{r}^\prime, t^\prime) \hat{E}_s (\vec{r}^\prime, t^\prime) .
\end{align}
\end{widetext}
The three different fields contributing to $\Ei$ up to second order in the interaction Hamiltonian have a clear interpretation: $\Ei_\mathrm{vac}$ is the vacuum field also present without the nonlinear coupling or the laser pulses; $\Ei_\mathrm{s}$ is the field generated by the mixing of the laser pulse $\Ecli$ with the THz vacuum field $\hat{E}$; $\Ei_\mathrm{2}$ is the field generated by the mixing of laser pulse $i$ with the THz source radiation $\hat{E}_s$.

\subsection{Identifying Source and Vacuum-Field Contributions}

In this section we identify source and vacuum field contributions to the EOS signal using symmetric operator ordering. To do so, we solve Heisenbergs equations of motions for the EOS signal operator $\hat{G}_{\theta_1 \theta_2}$ up to second order in $\chi^{(2)}$ using symmetric operator ordering between $\hat{E}$ and $\hat{E}^{(i)}$. This calculation follows along very similar lines as finding $G^{(AB)} $ in Appendix~\ref{app:FermiCorr}. We thus skip the details and give the resulting expressions, which are just the equivalent of Eqs.~\eqref{eq:SymmOrdVFApp} and \eqref{eq:SymmOrdSRApp} in case of Fermi's two atoms setup. 

As $\Eo_\mathrm{vac}$ and $\Et_\mathrm{vac}$ are uncorrelated there is no contribution to the EOS signal in $0$th order in $\chi^{(2)}$. The lowest-order non-vanishing contributions are of second order in $\chi^{(2)}$.

We obtain
\begin{align} \label{eq:GFull}
 G_{\theta_1 \theta_2} =  G_{\theta_1 \theta_2}\big|_\mathrm{vac}  +  G_{\theta_1 \theta_2}\big|_\mathrm{s} 
\end{align}
Here, the vacuum field contribution to the EOS signal is given by
\begin{align} \label{eq:GVFGerneral}
 G_{\theta_1 \theta_2}\big|_\mathrm{vac} & =\frac{1}{2C} \langle \{ \hat{S}^{(1)}_\mathrm{s},\hat{S}^{(2)}_\mathrm{s} \} \rangle ,  
 \end{align}
with 
\begin{multline} \label{eq:Sivac}
\hat{S}^{(i)}_\mathrm{s} = 4\pi \epsilon_0 c n_\mathrm{c} \int \dif^2 r_\parallel \int_0^\infty \dif \omega \frac{1}{\hbar \omega} \\
\times  [P(\theta_i) \mathcal{E}^{(i)\ast}(\vec{r}_\parallel, \omega)  \Ei_\mathrm{s}(\vec{r}_\parallel, \omega) + \mathrm{h.c.}].
\end{multline}
The fields at position $\vec{r}_\parallel$ are understood as fields evaluated at position $\vec{r} = ( \vec{r}_\parallel, L/2)^T$, i.e., in the plane of the backside of the nonlinear crystal. $G_{\theta_1 \theta_2}\big|_\mathrm{vac}$ probes correlations between $\hat{E}_\mathrm{s}^{(1)}$ and $\hat{E}_\mathrm{s}^{(2)}$. As $\hat{E}_\mathrm{s}^{(1)}$ and $\hat{E}_\mathrm{s}^{(2)}$ are generated via the individual mixing of each of the two uncorrelated laser pulses with the quantum vacuum, respectively, correlations between $\hat{E}_\mathrm{s}^{(1)}$ and $\hat{E}_\mathrm{s}^{(2)}$ only arise due to correlations present in the quantum vacuum. 

The source radiation contribution in Eq.~\eqref{eq:GFull} is given by  
 \begin{align} \label{eq:GSRGerneral}
 G_{\theta_1 \theta_2}\big|_\mathrm{s} & =\frac{1}{2C} \left( \langle \{ \hat{S}^{(1)}_\mathrm{vac},\hat{S}^{(2)}_2 \} \rangle+ \langle \{ \hat{S}^{(1)}_2,\hat{S}^{(2)}_\mathrm{vac} \} \rangle\right) , 
 \end{align}
with 
\begin{multline} \label{eq:SiSR}
\hat{S}^{(i)}_2 = 4\pi \epsilon_0 c n_\mathrm{c} \int \dif^2 r_\parallel \int_0^\infty \dif \omega \frac{1}{\hbar \omega} \\
\times  [P(\theta_i) \mathcal{E}^{(1)\ast}(\vec{r}_\parallel, \omega)  \Ei_2(\vec{r}_\parallel, \omega) + \mathrm{h.c.}],
\end{multline}
and 
\begin{multline}
\hat{S}^{(i)}_\mathrm{vac} = 4\pi \epsilon_0 c n_\mathrm{c} \int \dif^2 r_\parallel \int_0^\infty \dif \omega \frac{1}{\hbar \omega} \\
\times  [P(\theta_i) \mathcal{E}^{(1)\ast}(\vec{r}_\parallel, \omega)  \Ei_\mathrm{vac}(\vec{r}_\parallel, \omega) + \mathrm{h.c.}].
\end{multline}
$\Ei_\mathrm{2}$ given in Eq.~\eqref{eq:EOMTimeNIR2} is the field generated by the mixing of laser pulse $i$ with the THz source radiation $\hat{E}_s$. $\hat{E}_s$ can in turn be generated either by the mixing of $\Eclo$ with $\Eo_\mathrm{vac}$ or $\Eclt$ with $\Et_\mathrm{vac}$. Only the process in which laser pulse $1$ ($2$) interacts with the source radiation generated by the mixing of $\Eclt$ with $\Et_\mathrm{vac}$ (of $\Eclo$ with $\Eo_\mathrm{vac}$) leads to correlations between the two modes $\Eo$ and $\Et$, and, thus, to a contribution to the EOS signal. In this case $\Eo_\mathrm{2}$ ($\Et_\mathrm{2}$) is correlated with the NIR vacuum field in the other mode $\Et_\mathrm{vac}$ ($\Eo_\mathrm{vac}$) and thus with the shot noise contribution in the other detector.

Note that if we would have used a different operator ordering, we would have obtained a different splitting of the EOS signal into source and vacuum-field contributions. In the following we find explicit expressions for the vacuum-field and source-radiation contributions in Eqs.~\eqref{eq:GVFGerneral} and \eqref{eq:GSRGerneral}, respectively.

\subsection{Vacuum-Field Contribution}

To evaluate Eq.~\eqref{eq:GVFGerneral} we first insert the first order solution for $\hat{E}^{(i)}_{s}(\vec{r}, \omega)$ in Eq.~\eqref{eq:LikeDysonNIR} into the expression for $\hat{S}^{(i)}_\mathrm{s}$ in Eq.~\eqref{eq:Sivac} and find
\begin{widetext}
\begin{multline}
\hat{S}_\mathrm{s}^{(i)} = -8\pi \epsilon_0 c n_\mathrm{c} \chi^{(2)} \mu_0 \int \dif^2 r_\parallel \int_0^\infty \dif \omega \frac{\omega}{\hbar}  \int_{V_C} \dif ^3 r^\prime \int_{-\infty}^{\infty} \dif \Omega  \\
\times [P(\theta_i) \mathcal{E}^{(i)\ast}(\vec{r}_\parallel, \omega) \mathsf{D}(\vec{r}_\parallel, \vec{r}^\prime, \omega) \,\Ecli(\vec{r}^\prime, \omega-\Omega) \hat{E}_{\mathrm{vac}}(\vec{r}^\prime,\Omega)  + \mathrm{h.c.}].
\end{multline}
To simplify this expression we first insert the Green tensor in the paraxial approximation found in Eq.~\eqref{eq:GreensParaxial} and use
\begin{align} \label{eq:GroupRef}
k(\omega) \approx \frac{n(\omega_c) }{c}  \omega_c + \frac{n_{g}}{c} (\omega- \omega_c) ,
\end{align}
which gives
\begin{multline} \label{eq:S1General}
S_\mathrm{s}^{(i)} = - \frac{\sqrt{C}}{2L} \int_0^\infty \dif \omega  \int_{V_C} \dif ^3 r^\prime g_i^2(\vec{r}_\parallel^\prime) \int_{0}^{\infty} \dif \Omega \big\{ \hat{E}_{\mathrm{vac}} (\vec{r}^\prime,\Omega)   \underbrace{ \me^{\mi \Omega \delta t_i - \mi n_g \Omega z^\prime/c}  [ \mi P(\theta_i)f(-\Omega)    - \mi P^\ast(\theta_i) f(\Omega) ] }_{\equiv A_i(\Omega, z^\prime)} \\
+\hat{E}_{\mathrm{vac}}^{ \dagger} (\vec{r}^\prime,\Omega)  \underbrace{\me^{-\mi \Omega \delta t_i + \mi n_g \Omega z^\prime/c} [ \mi P(\theta_i) f(\Omega)    - \mi P^\ast(\theta_i) f(-\Omega) ] }_{\equiv A_i(-\Omega, z^\prime)}\big\}.
\end{multline}
Here, we also defined the total number of detected photons $N$, the average detected frequency $\omega_p$, the normalized spectral auto-correlation function $f(\Omega)$, and the detector efficiency $C$ \cite{benea-chelmus_electric_2019} via
\begin{align}\label{eq:Detector}
\begin{array}{llll}
\omega_p &=  \frac{\int_0^\infty \dif \omega   \mathcal{E}^{(i)2}(\omega)}{\int_0^\infty \dif \omega  \frac{1}{\omega} \mathcal{E}^{(i)2}(\omega)},  
 &  N & = 4\pi \epsilon_0 c n_\mathrm{c} \int_0^\infty \dif \omega  \frac{1}{\hbar \omega} \mathcal{E}^{(i)2}(\omega) ,\\
f(\Omega) & =  \frac{\int_0^\infty \dif \omega  \mathcal{E}^{(i)}(\omega)\mathcal{E}^{(i)}(\omega+ \Omega)}{\int_0^\infty \dif \omega   \mathcal{E}^{(i)2}(\omega)},  
\quad \quad  & \sqrt{C} &= \frac{2 L \chi^{(2)} N \omega_p}{\epsilon_0 c n_\mathrm{c}}.
\end{array}
\end{align}
Inserting Eq.~\eqref{eq:S1General} into Eq.~\eqref{eq:GVFGerneral} for both $\hat{S}^{(1)}_\mathrm{s}$ and $\hat{S}^{(2)}_\mathrm{s}$ we find the EOS signal stemming from vacuum fluctuations
\begin{multline}
G_{\theta_1 \theta_2}\big|_\mathrm{vac} =\frac{1 }{8L^2}   \int_{V_C} \dif ^3 r^\prime \int_{V_C} \dif ^3 r^{\prime\prime}  \int_{0}^{\infty} \dif \Omega  \int_{0}^{\infty} \dif \Omega^\prime  \, \langle \hat{E}_{\mathrm{vac}} (\vec{r}^\prime,\Omega) \hat{E}_{\mathrm{vac}}^\dagger (\vec{r}^{\prime\prime},\Omega^\prime) \rangle \\
\times \left\{ g_1^2(\vec{r}_\parallel^\prime) g_2^2(\vec{r}_\parallel^{\prime\prime}) A_1(\Omega, z^\prime) A_2(-\Omega^\prime, z^{\prime\prime})+ g_2^2(\vec{r}_\parallel^\prime) g_1^2(\vec{r}_\parallel^{\prime\prime})   A_1(-\Omega^\prime, z^{\prime\prime})  A_2(\Omega, z^{\prime}) \right\}  .
\end{multline}
We can take the vacuum expectation value using Eq.~\eqref{eq:EvacEvactoGreens}, use the Onsager reciprocity relation $\mathsf{D}(\vec{r}^\prime, \vec{r}, \Omega) = \mathsf{D}(\vec{r}, \vec{r}^\prime, \Omega)$ \cite{buhmann2013dispersion} 
, and Schwartz reflection principle, i.e., $\mathsf{D}(-\Omega) = \mathsf{D}^\ast(\Omega)$ \cite{buhmann2013dispersion} to find
\begin{align} \label{eq:GVFFreqAlmost}
G_{\theta_1 \theta_2}\big|_\mathrm{vac}
 =\frac{1}{ 4 L^2 }   \int_{V_C} \dif ^3 r^\prime \int_{V_C} \dif ^3 r^{\prime\prime}  g_1^2(\vec{r}_\parallel^\prime) g_2^2(\vec{r}_\parallel^{\prime\prime}) \int_{-\infty}^{\infty} \dif \Omega   A_1(\Omega,z^\prime) A_2(-\Omega, z^{\prime\prime})\mathcal{C}(\vec{r}^\prime, \vec{r}^{\prime\prime}, \Omega) .
\end{align}
Here we have identified the correlation function in frequency domain $\mathcal{C}(\vec{r}^\prime, \vec{r}^{\prime\prime}, \Omega) =   \frac{\hbar \mu_0}{2\pi} \mathrm{sgn}[\Omega]\Omega^2\mathrm{Im}[\mathsf{D}(\vec{r}^\prime, \vec{r}^{\prime\prime}, \Omega)] $, see Eq.~\eqref{eq:CorrelationFunctionFrequency}. We use $f_1(\Omega) = f_2(\Omega) \equiv f(\Omega)$ and $f(-\Omega) = f(\Omega)$, and set $\delta t_1 = 0$, and $\delta t_2  = \delta t$, such that Eq.~\eqref{eq:GVFFreqAlmost} reduces to
\begin{align} \label{eq:GVFFreq}
G_{\theta_1 \theta_2}\big|_\mathrm{vac}
 = \mathrm{Im}[P(\theta_1)]\mathrm{Im}[P(\theta_2)]  \int_{V_C} \dif ^3 r^\prime \int_{V_C} \dif ^3 r^{\prime\prime}  \int_{-\infty}^{\infty} \dif \Omega  F(\vec{r}^\prime, \vec{r}^{\prime\prime}, \Omega)\mathcal{C}(\vec{r}^\prime, \vec{r}^{\prime\prime}, \Omega) ,
\end{align}
with 
\begin{align}
 F(\vec{r}^\prime, \vec{r}^{\prime\prime}, \Omega) = \frac{1}{L^2} g_1^2(\vec{r}_\parallel^\prime) g_2^2(\vec{r}_\parallel^{\prime\prime}) \me^{-\mi n_g \Omega(z^\prime-z^{\prime\prime})/c} \me^{-\mi \delta t \Omega} f^2(\Omega).
\end{align}
In case $P(\theta_1) = P(\theta_2) = \mi$ this result has been previously obtained in Refs.~\cite{lindel2020theory,lindel2021macroscopic}. In time domain, we use Eq.~\eqref{eq:GVFFreqAlmost} and 
\begin{subequations} \label{eq:Bis}
 \begin{align} 
 g_1^2(\vec{r}_\parallel) \int_{-\infty}^\infty \dif \Omega   \me^{-\mi \Omega t }A_1(\Omega, z^\prime)  & = - 4 \pi L\, L_1(\vec{r},t) \mathrm{Im} [  P(\theta_1) ], \\
 g_2^2(\vec{r}_\parallel) \int_{-\infty}^\infty \dif \Omega   \me^{\mi \Omega t }A_2(-\Omega, z^\prime) & = - 4 \pi L\, L_2(\vec{r},t) \mathrm{Im} [  P(\theta_2) ],
 \end{align}
 \end{subequations} 
 to find
 \begin{align} \label{eq:GVFTimeNoFilter}
G_{\theta_1 \theta_2}\big|_\mathrm{vac}
 =\mathrm{Im}[P(\theta_1)]\mathrm{Im}[P(\theta_2)]  \int_{V_C} \dif ^3 r^\prime\int_{V_C} \dif ^3 r^{\prime\prime} \int_{-\infty}^{\infty} \dif t  \int_{-\infty}^{\infty} \dif t^\prime L_{1}(\vec{r}^\prime, t) L_{2}(\vec{r}^{\prime\prime}, t^{\prime})\mathcal{C}(\Greektens{$\rho$},\tau). 
\end{align}
This is Eq.~(4) of the main text. Equation \eqref{eq:GVFTimeNoFilter} has been previously obtained in Ref.~\cite{settembrini2022detection} in case both modes are detected using quarter-wave plates so that $P(\theta_1) = P(\theta_2) = \mi$ in which case
 \begin{align} \label{eq:GVFFinal}
G_\mathrm{vac} \equiv G_{\frac{\pi}{2}\frac{\pi}{2}}\big|_\mathrm{vac}
 =\int_{V_C} \dif ^3 r^\prime\int_{V_C} \dif ^3 r^{\prime\prime} \int_{-\infty}^{\infty} \dif t  \int_{-\infty}^{\infty} \dif t^\prime L_{1}(\vec{r}^\prime, t) L_{2}(\vec{r}^{\prime\prime}, t^{\prime})\mathcal{C}(\Greektens{$\rho$},\tau). 
\end{align}

\subsection{Source-Radiation Contribution}

To evaluate the source radiation contribution to the EOS signal in Eq.~\eqref{eq:GSRGerneral}. We start with simplifying the expression for $\Ei_2$ by Fourier transforming Eq.~\eqref{eq:EOMTimeNIR2}:
\begin{align} \label{eq:EnirSource2} 
\hat{E}_{2}^{(i)}(\vec{r}, \omega) =- 2 \chi^{(2)} \mu_0 \omega^2 \int_{V_C} \dif ^3 r^\prime \int_{-\infty}^{\infty} \dif \Omega \mathsf{D}(\vec{r}, \vec{r}^\prime, \omega) \, \Ecli(\vec{r}^\prime, \omega-\Omega) \hat{E}_{\mathrm{s}} (\vec{r}^\prime,\Omega) .
\end{align}
Inserting Eq.~\eqref{eq:LikeDysonTHz} into Eq.~\eqref{eq:EnirSource2} we find 
\begin{multline} \label{eq:EnirSource3} 
\hat{E}_{2}^{(i)}(\vec{r}, \omega) = 4 \chi^{(2) 2} \mu_0^2 \omega^2 \int_{V_C} \dif ^3 r^\prime  \int_{V_C} \dif ^3 r^{\prime\prime}\int_{-\infty}^{\infty} \dif \Omega \, \Omega^2  \int_{0}^{\infty} \dif \omega^\prime  \mathsf{D}(\vec{r}, \vec{r}^\prime, \omega) \mathsf{D}(\vec{r}^\prime, \vec{r}^{\prime\prime}, \Omega) \, \Ecli(\vec{r}^\prime, \omega-\Omega) \\
\times \left[ \mathcal{E}^{(\bar{i})}(\vec{r}^{\prime\prime}, \Omega-\omega^\prime) \hat{E}^{(\bar{i})}_\mathrm{vac} (\vec{r}^{\prime\prime}, \omega^\prime)  + \mathcal{E}^{(\bar{i})}(\vec{r}^{\prime\prime}, \Omega+\omega^\prime)\hat{E}^{(\bar{i})\dagger}_\mathrm{vac} (\vec{r}^{\prime\prime}, \omega^\prime) \right].
\end{multline}
Here, $\bar{i} = 2,1$ if $i = 1,2$. Note that we have only included source terms of the THz field stemming from the other laser pulse, since otherwise the expectation value in Eq.~\eqref{eq:GSRGerneral} would vanish as $\langle \hat{E}_{i} \hat{E}_{\bar{i}} \rangle = 0$. Physically, this is also clear: to generate correlations between $\hat{E}^{(i)} $ and $\hat{E}^{(\bar{i})} $, we have to consider the influence of the source radiation from mode $i$ onto $\hat{E}^{(\bar{i})}$ and vice versa. Eventually, we need to evaluate $\left\{ \hat{S}^{(i)}_2, \hat{S}^{(\bar{i})}_\mathrm{vac} \right\}$ which can be written in terms of $\Ei_2$ as 
\begin{multline} \label{eq:SRVeryearlystage}
\left\{ \hat{S}^{(i)}_2, \hat{S}^{(\bar{i})}_\mathrm{vac} \right\} = 16 \pi^2 \epsilon_0^2 c^2 n_\mathrm{c}^2 \int \dif^2 r_\parallel \int \dif^2 r_\parallel^\prime \int_0^{\infty} \dif \omega \int_0^{\infty} \dif \omega^\prime \frac{1}{\hbar^2 \omega \omega^\prime}  \\
\times \left( P(\theta_{\bar{i}}) \mathcal{E}^{(\bar{i})\ast}(\vec{r}_\parallel^\prime, \omega^\prime)  \hat{E}^{(\bar{i})}_\mathrm{vac}(\vec{r}_\parallel^\prime, \omega^\prime)  [P(\theta_i) \mathcal{E}^{(i)\ast}(\vec{r}_\parallel, \omega)  \Ei_2(\vec{r}_\parallel, \omega) + \mathrm{h.c.}] \right. \\
\left.+   [P(\theta_i) \mathcal{E}^{(i)\ast}(\vec{r}_\parallel, \omega)  \Ei_2(\vec{r}_\parallel, \omega) + \mathrm{h.c.}] P^\ast(\theta_{\bar{i}}) \mathcal{E}^{(\bar{i})}(\vec{r}_\parallel^\prime, \omega^\prime)  \hat{E}^{(\bar{i})\dagger}_\mathrm{vac}(\vec{r}_\parallel^\prime, \omega^\prime) \right) ,
\end{multline}
where we also used $  \Ei_\mathrm{vac} |0\rangle = 0$ and $\langle 0 |  \hat{E}^{(i)\dagger}_\mathrm{vac} = 0$. We thus encounter terms of the form $\langle \hat{E}_{2}^{(i)}(\vec{r}_\parallel, \omega) \hat{E}^{(\bar{i})\dagger}_\mathrm{vac}(\overline{\vec{r}_\parallel}, \overline{\omega}) \rangle $ into which we insert Eq.~\eqref{eq:EnirSource3} and use Eq.~\eqref{eq:EvacEvactoGreens} to find
\begin{multline}       
\langle \hat{E}_{2}^{(i)}(\vec{r}_\parallel, \omega) \hat{E}^{(\bar{i})\dagger}_\mathrm{vac}(\overline{\vec{r}_\parallel}, \overline{\omega}) \rangle  =  \frac{4 \chi^{(2) 2} \hbar \mu_0^3 \omega^2 \overline{\omega}^2}{\pi} \int_{V_C} \dif ^3 r^\prime  \int_{V_C} \dif ^3 r^{\prime\prime}\int_{-\infty}^{\infty} \dif \Omega \, \Omega^2   \mathsf{D}(\vec{r}_\parallel, \vec{r}^\prime, \omega) \\
 \quad \quad \times \mathsf{D}(\vec{r}^\prime, \vec{r}^{\prime\prime}, \Omega) \, \Ecli(\vec{r}^\prime, \omega-\Omega)  \mathcal{E}^{(\bar{i})}(\vec{r}^{\prime\prime}, \Omega-\overline{\omega})\mathrm{Im} \mathsf{D}(\vec{r}^{\prime\prime}, \overline{\vec{r}_\parallel}, \overline{\omega}).
\end{multline}
We use the paraxial form of the Green tensors in the NIR frequency range in Eq.~\eqref{eq:GreensParaxial} to further simplify this expression:
\begin{multline} \label{eq:Help1}
\langle \hat{E}_{2}^{(i)}(\vec{r}_\parallel, \omega) \hat{E}^{(\bar{i})\dagger}_\mathrm{vac}(\overline{\vec{r}}_\parallel, \overline{\omega}) \rangle 
 = \frac{ \mi \chi^{(2) 2} \hbar \mu_0^3 \omega \overline{\omega} c^2}{2\pi n_\mathrm{c}^2} \int_{-L/2}^{L/2} \dif z^\prime  \int_{-L/2}^{L/2} \dif z^{\prime\prime}\int_{-\infty}^{\infty} \dif \Omega \, \Omega^2  \Ecli (\omega-\Omega) \mathcal{E}^{(\bar{i})}(\Omega -\overline{\omega}) g_{i}(\vec{r}_\parallel) g_{\bar{i}}(\overline{\vec{r}}_\parallel) \\
\times \me^{-\mi n_g \Omega (z^{\prime}- z^{\prime\prime})/c} \mathsf{D}(\{\vec{r}_\parallel, z^\prime \}, \{\overline{\vec{r}}_\parallel, z^{\prime\prime} \}, \Omega).
\end{multline}
Similar expressions can be obtained for the other terms in Eq.~\eqref{eq:SRVeryearlystage} and we eventually find
\begin{multline} \label{eq:S2Complicated7}
\big \langle \left\{ \hat{S}^{(i)}_2, \hat{S}^{(\bar{i})}_\mathrm{vac} \right\} \big \rangle  = \frac{ 2C \hbar \mu_0 }{16L^2 \pi}\int_{V_C} \dif^3 r^\prime  \int_{V_C} \dif^3 r^{\prime\prime}   g_{i}^2(\vec{r}_\parallel^\prime) g^{2}_{\bar{i}}(\vec{r}_\parallel^{\prime\prime})   \int_{-\infty}^\infty \dif \Omega \, \Omega^2 \me^{-\mi n_g \Omega (z^\prime-z^{\prime\prime})/c+ \mi (\delta t_i - \delta t_{\bar{i}})\Omega} \mathsf{D}(\vec{r}^{\prime}, \vec{r}^{\prime\prime}, \Omega)   \\
\times    \mi \left[  P(\theta_i)P^\ast(\theta_{\bar{i}}) f^2(-\Omega)-  P^{\ast }(\theta_i)  P^\ast(\theta_{\bar{i}})f(\Omega)f(-\Omega) - P^\ast(\theta_i) P(\theta_{\bar{i}}) f^2(\Omega)  +  P(\theta_i) P(\theta_{\bar{i}}) f(\Omega) f(-\Omega)\right] .
\end{multline}
Using $ \mathcal{R}(\vec{r}^\prime, \vec{r}^{\prime\prime}, \Omega)  = \Omega^2 \frac{ \mu_0}{2\pi} \mathsf{D}(\vec{r}^{\prime}, \vec{r}^{\prime\prime}, \Omega) $ as well as $f(-\Omega) = f(\Omega) $ and $f_1(\Omega) = f_2(\Omega) \equiv f(\Omega)$, Eq.~\eqref{eq:S2Complicated7} simplifies to 
\begin{multline} \label{eq:S2Complicated8}
\big \langle \left\{ \hat{S}^{(i)}_2, \hat{S}^{(\bar{i})}_\mathrm{vac} \right\} \big \rangle  = -\frac{ 2C \hbar }{4L^2 }\int_{V_C} \dif^3 r^\prime  \int_{V_C} \dif^3 r^{\prime\prime}   g_{i}^2(\vec{r}_\parallel^\prime) g^{2}_{\bar{i}}(\vec{r}_\parallel^{\prime\prime})   \int_{-\infty}^\infty \dif \Omega \, f^2(\Omega) \me^{-\mi n_g \Omega (z^\prime-z^{\prime\prime})/c+ \mi (\delta t_i - \delta t_{\bar{i}})\Omega}  \mathcal{R}(\vec{r}^{\prime}, \vec{r}^{\prime\prime}, \Omega)   \\
\times    \mathrm{Im} \left[  P(\theta_i)P^\ast(\theta_{\bar{i}})  + P(\theta_i) P(\theta_{\bar{i}}) \right] .
\end{multline}
Summing over $i$ and multiplying by $1/2C$, we find the source radiation contribution to the EOS signal [compare Eq.~\eqref{eq:GSRGerneral}]
\begin{multline} \label{eq:GRSRSimplified}
G_{\theta_1 \theta_2}\big|_\mathrm{s} = -\frac{ \hbar }{2L^2 }\int_{V_C} \dif^3 r^\prime  \int_{V_C} \dif^3 r^{\prime\prime}   g_{1}^2(\vec{r}_\parallel^\prime) g^{2}_{2}(\vec{r}_\parallel^{\prime\prime})   \int_{-\infty}^\infty \dif \Omega \,  f^2(\Omega) \me^{\mi \Phi \Omega} \\
\times  \left\{\mi \mathrm{Im}[\mathcal{R}(\vec{r}^{\prime}, \vec{r}^{\prime\prime}, \Omega)]\mathrm{Im} \left[  P(\theta_1)P^\ast(\theta_{2}) \right]   + \mathrm{Re}[\mathcal{R}(\vec{r}^{\prime}, \vec{r}^{\prime\prime}, \Omega)]  \mathrm{Im} \left[ P(\theta_1) P(\theta_{2}) \right] \right\}.
\end{multline}
Here, we have defined $\Phi = - n_g (z^\prime-z^{\prime\prime})/c+  (\delta t_1 - \delta t_{2})$.
Shifting to positive frequencies, we get
\begin{multline} \label{eq:GRSRSimplified05}
G_{\theta_1 \theta_2}\big|_\mathrm{s}  = \frac{ \hbar }{ L^2 }\int_{V_C} \dif^3 r^\prime  \int_{V_C} \dif^3 r^{\prime\prime}   g_{1}^2(\vec{r}_\parallel^\prime) g^{2}_{2}(\vec{r}_\parallel^{\prime\prime})   \int_{0}^\infty \dif \Omega \,  f^2(\Omega)  \\
\times  \left\{ \mathrm{Im}[\mathcal{R}(\vec{r}^{\prime}, \vec{r}^{\prime\prime}, \Omega)] \mathrm{sin}[\Phi \Omega] \mathrm{Im} \left[  P(\theta_1)P^\ast(\theta_{2}) \right]   - \mathrm{Re}[\mathcal{R}(\vec{r}^{\prime}, \vec{r}^{\prime\prime}, \Omega)]  \mathrm{cos}[\Phi \Omega] \mathrm{Im} \left[ P(\theta_1) P(\theta_{2}) \right] \right\}.
\end{multline}
Note that $\mathrm{Re}[\mathcal{R}(\Omega)]  = \mathcal{R}^\prime (\Omega)$ and $ \mi \mathrm{Im}[\mathcal{R}(\Omega)]  =  \mathcal{R}^{\prime\prime} (\Omega)$. To find the source radiation contribution to the EOS signal as a function of the time-domain response function, we Fourier transform the reactive and dissipative part of the response function and find
\begin{align}   \label{eq:GSRTimeSecond2}
G_{\theta_1 \theta_2}\big|_\mathrm{s}
& = -\frac{\hbar}{2} \int_{\vec{r}, \vec{r}^\prime, t, t^\prime }   L_1(\vec{r}, t ) L_2(\vec{r}^\prime, t^\prime ) \left\{  \mathrm{Im}\left[  P(\theta_1) P(\theta_2)  \right]  \mathcal{R}^\prime(\Greektens{$\rho$}, \tau) +  \mathrm{Im}\left[  P(\theta_1) P^\ast(\theta_2) \right] \mathcal{R}^{\prime\prime}(\Greektens{$\rho$}, \tau)          \right\} 
\end{align}
This is Eq.~(7) in the main text.

For a quarter and half wave plate for mode $1$ and $2$, respectively, we find $P(\theta_1 = \pi/2) = \mi$ and $P(\theta_2 = \pi) = 1$ such that 
\begin{align} \label{eq:GSRTimeSecond}
G_{\frac{\pi}{2} \pi}\big|_\mathrm{s} \equiv  G^{(1)}_{s}  = - \frac{\hbar}{2} \int_{\vec{r}, \vec{r}^\prime, t, t^\prime }   L_1(\vec{r}, t ) L_2(\vec{r}^\prime, t^\prime ) \left\{\mathcal{R}^\prime(\Greektens{$\rho$}, \tau) +   \mathcal{R}^{\prime\prime}(\Greektens{$\rho$}, \tau)          \right\} = - \frac{\hbar}{2} \int_{\vec{r}, \vec{r}^\prime, t, t^\prime }   L_1(\vec{r}, t ) L_2(\vec{r}^\prime, t^\prime ) \mathcal{R}(\Greektens{$\rho$}, \tau) .
\end{align}
This is Eq.~(8) of the main text. We see that for this arrangement of the detection scheme the setup is sensitive to detect source radiation propagating from mode $2$ to mode $1$. \\
Furthermore, we find $P(2\pi/3) = (1+\mi)/\sqrt{2} = P^\ast(4\pi/3)$ such that
\begin{align}
G_{\mathcal{R}^{\prime}} & \equiv \frac{1}{2}\left(  G_{\frac{2\pi}{3} \frac{2\pi}{3}}-G_{\frac{4\pi}{3}\frac{4\pi}{3}}\right)  =   - \frac{\hbar}{2} \int_{\vec{r}, \vec{r}^\prime, t, t^\prime }   L_1(\vec{r}, t ) L_2(\vec{r}^\prime, t^\prime ) \mathcal{R}^{\prime}(\Greektens{$\rho$}, \tau), \\
G_{\mathcal{R}^{\prime \prime}} &  \equiv \frac{1}{2}\left(  G_{\frac{\pi}{2} \pi}-G_{\pi \frac{\pi}{2}}\right)  =   - \frac{\hbar}{2} \int_{\vec{r}, \vec{r}^\prime, t, t^\prime }   L_1(\vec{r}, t ) L_2(\vec{r}^\prime, t^\prime ) \mathcal{R}^{\prime \prime }(\Greektens{$\rho$}, \tau).
\end{align}

\subsection{Fluctuation--Dissipation Theorem} \label{sec:FDTApp}

The implication of the time-domain fluctuation--dissipation theorem in Eq.~\eqref{eq:TFDT} onto the EOS signal can be found by calculating the Hilbert transform of $G_{\mathcal{R}^{\prime\prime}}$ with respect to $\delta t$:
\begin{subequations} \label{eq:EOSTFDTDerivation}
\begin{align}
\mathcal{H} G_{\mathcal{R}^{\prime\prime}}(\delta t) &= - \frac{\hbar}{2} \mathcal{H}\int_{\vec{r}, \vec{r}^\prime,t, t^\prime} L_1(\vec{r}, t) L_2(\vec{r}^\prime, t^\prime )  \mathcal{R}^{\prime\prime}(\boldsymbol{\rho}, \tau) \\
 &= - \frac{\hbar}{2} \int_{\vec{r}, \vec{r}^\prime,t, t^\prime} L_1(\vec{r}, t) L_2(\vec{r}^\prime, t^\prime - \delta t) \mathcal{H}\mathcal{R}^{\prime\prime}(\boldsymbol{\rho}, \tau +\delta t) \\
 &=   \frac{1}{2} \int_{\vec{r}, \vec{r}^\prime,t, t^\prime} L_1(\vec{r}, t) L_2(\vec{r}^\prime, t^\prime ) \mathcal{C}(\boldsymbol{\rho}, \tau )  =  \frac{1}{2} G_\mathrm{vac}(\delta t)
\end{align}
\end{subequations}
Note that  $L_2(\vec{r}^\prime, t^\prime - \delta t)$ is independent of $\delta t$ and in the last line we used the time-domain fluctuation--dissipation theorem in Eq.~\eqref{eq:TFDT}. Equation~\eqref{eq:EOSTFDTDerivation} gives Eq.~\eqref{eq:EOSTFDT} of the main text. \\
By Fourier transforming the different contributions to the EOS signal with respect to $\delta t$ one obtains the frequency domain EOS signals (remember $G_i(\Omega) \equiv \frac{1}{2\pi} \int_{-\infty}^\infty \dif \delta t \, \me^{\mi \Omega \delta t }G_i(\delta t)  $, with $i = \mathrm{vac}, \mathrm{s}, \mathcal{R}^{\prime \prime}$)
\begin{align}
G_{\mathrm{vac}}(\Omega)   &  =\frac{1}{L^2}  \int_{V_C} \dif ^3 r^\prime\int_{V_C} \dif ^3 r^{\prime\prime}   g_1^2(\vec{r}_\parallel^\prime) g_2^2(\vec{r}_\parallel^{\prime\prime})  f^2(\Omega)  \me^{-\mi n_g \Omega(z^\prime-z^{\prime\prime})/c}  \mathcal{C}(\vec{r}^\prime, \vec{r}^{\prime\prime}, \Omega) ,\\
G_{\mathrm{s}}(\Omega)   &  =  -\frac{ \hbar }{2L^2 }\int_{V_C} \dif^3 r^\prime  \int_{V_C} \dif^3 r^{\prime\prime}   g_{1}^2(\vec{r}_\parallel^\prime) g^{2}_{2}(\vec{r}_\parallel^{\prime\prime})   f^2(\Omega) \me^{-\mi n_g \Omega(z^\prime-z^{\prime\prime})/c}  \mathcal{R}(\vec{r}^\prime, \vec{r}^{\prime\prime}, \Omega) , \\      \label{eq:Last}
 G_{\mathcal{R}^{\prime\prime}}(\Omega)   & =  -\frac{ \hbar }{2L^2 }\int_{V_C} \dif^3 r^\prime  \int_{V_C} \dif^3 r^{\prime\prime}   g_{1}^2(\vec{r}_\parallel^\prime) g^{2}_{2}(\vec{r}_\parallel^{\prime\prime})   f^2(\Omega) \me^{-\mi n_g \Omega(z^\prime-z^{\prime\prime})/c}  \mathcal{R}^{\prime\prime}(\vec{r}^\prime, \vec{r}^{\prime\prime}, \Omega) 
 =\mi \mathrm{Im} G_{\mathrm{s}}(\Omega).
\end{align}
These expressions can be verified by using Eqs.~\eqref{eq:GVFFreq} and \eqref{eq:GRSRSimplified} for $G_\mathrm{vac}$ and $G_{\mathrm{s}} $, respectively. The last equality sign in Eq.~\eqref{eq:Last} can be seen from $\mi \mathrm{Im}\mathcal{R}(\Omega) = \mathcal{R}^{\prime\prime}(\Omega)$.

\subsection{Energy Conservation.}

Here, we add a discussion on the issue of energy conservation in EOS experiments. In close analogy to the stability of ground-state atoms and energy conservation in the atom's dynamics, where the loss of energy due to the emission of source radiation by the fluctuating charges is canceled by the process in which the atom absorbs energy from the vacuum \cite{milonni1994quantum}, we find that the individual source radiation and vacuum field fluctuation contributions include terms in which energy is extracted from the quantum vacuum. These contributions only cancel when adding the two contributions. \\

To see this, we sum over $i$ in Eq.~\eqref{eq:S2Complicated7} to find the source radiation contribution in its most general form (not assuming that the spectral autocorrelation function $f(\Omega)$ is symmetric)
\begin{multline} \label{eq:EnergyConservSR}
G_{\theta_1\theta_2}\big|_\mathrm{s}= \frac{ \hbar \mu_0  }{4 L^2 \pi }\int_{V_C} \dif^3 r^\prime  \int_{V_C} \dif^3 r^{\prime\prime}   g_{1}^2(\vec{r}_\parallel^\prime) g^{2}_{2}(\vec{r}_\parallel^{\prime\prime})   \int_{0}^\infty \dif \Omega \, \Omega^2   \\ 
\times  \left\{  \mathrm{Im}[\mathsf{D}(\vec{r}^{\prime}, \vec{r}^{\prime\prime}, \Omega)] \left(f^2(\Omega)   \mathrm{Re}[ \me^{\mi \Phi \Omega}  P^\ast(\theta_1) P(\theta_{2}) ] - f^2(-\Omega) \mathrm{Re}[ \me^{-\mi \Phi \Omega}  P^\ast(\theta_1) P(\theta_{2}) ]  \right) \right.   \\
 \left. -  \mathrm{Re}[\mathsf{D}(\vec{r}^{\prime}, \vec{r}^{\prime\prime}, \Omega)] f(\Omega)f(-\Omega)  \mathrm{Im}[ \me^{\mi \Phi \Omega}  P(\theta_1) P(\theta_{2})+ \me^{-\mi \Phi \Omega}  P(\theta_1) P(\theta_{2})] \right\}.
\end{multline}
This expression can be compared to the vacuum field contribution in its most general form, which can be obtained from Eq.~\eqref{eq:GVFFreqAlmost}:
\begin{multline} \label{eq:GVFFull}
G_{\theta_1\theta_2}\big|_\mathrm{vac}
 =\frac{\hbar \mu_0 }{4L^2 \pi}   \int_{V_C} \dif ^3 r^\prime\int_{V_C} \dif ^3 r^{\prime\prime}  g_1^{2}(\vec{r}_\parallel^\prime) g_2^{ 2}(\vec{r}_\parallel^{\prime\prime}) \int_{0}^{\infty} \dif \Omega   \, \Omega^2 \mathrm{Im}[\mathsf{D}(\vec{r}^\prime, \vec{r}^{\prime\prime}, \Omega)] \\
\times \left\{ f^2(\Omega) \mathrm{Re}[\me^{\mi \Omega \phi}P^\ast(\theta_1)P(\theta_2)] + f^2(-\Omega) \mathrm{Re}[\me^{-\mi \Omega \phi}P^\ast(\theta_1)P(\theta_2) ] \right. \\
\left. - 2 f(-\Omega)f(\Omega) \mathrm{Re}[P(\theta_1)P(\theta_2)] \mathrm{cos}[\Phi \Omega]\right\}.
\end{multline}
In the source-radiation and in the vacuum-field contribution to the signal in Eqs.~\eqref{eq:EnergyConservSR} and \eqref{eq:GVFFull}, respectively, we find energy non-conserving contributions which are proportional to $f^2(-\Omega)$. As $f(-\Omega) \propto \mathcal{E}(\omega- \Omega)$. These terms correspond to generating two photons with energy $\hbar \omega $ by absorbing photons from the laser pulse which have the lower energy $\hbar(\omega -\Omega)$, e.g., by sum-frequency generation of a laser and a vacuum photon. These terms lead to non-vanishing contributions to $G_{\theta_1\theta_2}\big|_\mathrm{vac}$ and $G_{\theta_1\theta_2}\big|_\mathrm{s}$. However, when adding both contributions we find that these terms cancel for any values of $\theta_1$ and $\theta_2$, such that there are no energy non-conserving terms contributing to the experimental accessible full EOS signal $G_{\theta_1\theta_2}$, which is given by adding Eqs.~\eqref{eq:GVFFull} and \eqref{eq:EnergyConservSR} 
\begin{multline} \label{eq:EnergyConservFull}
G_{\theta_1\theta_2}= \frac{ \hbar \mu_0  }{2 L^2 \pi }\int_{V_C} \dif^3 r^\prime  \int_{V_C} \dif^3 r^{\prime\prime}   g_{1}^2(\vec{r}_\parallel^\prime) g^{2}_{2}(\vec{r}_\parallel^{\prime\prime})   \int_{0}^\infty \dif \Omega \, \Omega^2     \left\{  \mathrm{Im}[\mathsf{D}(\vec{r}^{\prime}, \vec{r}^{\prime\prime}, \Omega)] f^2(\Omega)   \mathrm{Re}[ \me^{\mi \Phi \Omega}  P^\ast(\theta_1) P(\theta_{2}) ]  \right.   \\
 \left. - f(\Omega)f(-\Omega) \mathrm{cos}[\Phi \Omega]  \mathrm{Im}[ P(\theta_1) P(\theta_{2}) \mathsf{D}(\vec{r}^{\prime}, \vec{r}^{\prime\prime}, \Omega) ] \right\}.
\end{multline}
This illustrates once more the necessity of including both `sides of the same quantum-mechanical coin' \cite{senitzky_radiation-reaction_1973}, vacuum field fluctuation and source radiation, to ensure energy conservation. It is in close analogy to the stability of ground state atoms in vacuum, which arises because the process where the atom is excited by a vacuum photon is compensated for by the energy loss due to the emission of source radiation of the fluctuating charges in the ground-state atom \cite{milonni1994quantum}.

\subsection{Rectangular Pulses} \label{app:EOSRec}

In this section, we derive simplified expressions for the EOS signal stemming from source radiation $G_\mathrm{s}$ and from vacuum field fluctuations $G_\mathrm{vac}$ in case of negligible absorption and dispersion effects in the crystal, i.e., we assume that $n(\Omega) \approx n \in \mathbb{R}$. We further assume that the laser pulses have a rectangular shape given by   
\begin{align} \label{eq:RectangularPulse}
L_1 (\vec{r}, t) = L_2(\vec{r}- \delta \vec{r}_\parallel, t-\delta t ) \frac{1}{L \tau_\sigma w^2} H\left[ \frac{x}{w}\right]  H\left[ \frac{y}{w}\right] H\left[ \frac{t- n_g z/c}{\tau_\mathrm{p}}\right].
\end{align}
Here, $ H[x]$ is a rectangular function, i.e., $H[x] = 1$ if $x \in [-0.5,0.5]$ and $H[x] = 0$ otherwise. The results found in this section have been used to generate the solid lines in Fig.~\ref{fig:EOSSignalDelt} of the main text. 

\paragraph{Source Radiation Contribution.}

Since we assume the refractive index to be constant in this section we can use the response function in Eq.~\eqref{eq:ResponseNormalMode}. Inserting Eq.~\eqref{eq:ResponseNormalMode} into Eq.~\eqref{eq:GSRTimeSecond} we find
\begin{align} \label{eq:GSR1}
G_\mathrm{s} = - \frac{ \hbar}{8 \pi \epsilon_0 c n}\int \dif t \int_{V_C} \!\!\! \dif^3 r \int \dif t^\prime \int_{V_C} \!\!\!  \dif^3  r^\prime  L_1(\vec{r}, t) L_2(\vec{r}^\prime, t^\prime)      \left(\frac{\partial^2}{\partial t \partial t^\prime} - c_n^2 \frac{\partial^2}{\partial x \partial x^\prime}\right) \frac{1}{\rho}  \delta(\rho- c_n \tau) .
\end{align}
Next, we insert Eq.~\eqref{eq:RectangularPulse} into Eq.\eqref{eq:GSR1}, use integration by parts to shift the derivatives to the pulse envelopes and substitute the integration variables to $\tau = t-t^\prime$ and $\Greektens{$\rho$} = \vec{r}- \vec{r}^\prime$ and obtain
 \begin{multline} \label{eq:GSR3}
G_\mathrm{s} =- \frac{ \hbar}{8  L w^2 \tau_\mathrm{p}^2 \pi \epsilon_0 c n } \left\{  \int \dif^3 \rho    \Lambda \left[ \frac{\rho_x - \delta x}{w}  \right] \Lambda\left[ \frac{\rho_y - \delta y}{w}  \right] \Lambda \left[ \frac{\rho_z}{L}\right]    \frac{1}{\rho} \delta(\rho-  c_n \tau) \bigg|_{t=n_g \frac{z}{c} -\frac{\tau_\mathrm{p}}{2}}^{t=n_g \frac{z}{c} +\frac{\tau_\mathrm{p}}{2}} \bigg|_{t^\prime =n_g \frac{z^\prime}{c}- \delta t -\frac{\tau_\mathrm{p}}{2}}^{t^\prime=n_g \frac{z^\prime}{c} - \delta t +\frac{\tau_\mathrm{p}}{2}} \right. \\
\left.   - \frac{\tau_\mathrm{p}  c_n}{w}  \int \dif \rho_z \int \dif \rho_y  \Lambda \left[ \frac{\rho_z}{L}\right]  \frac{ \Lambda\left[ \frac{\rho_y - \delta y}{w}  \right]  \Lambda \left[\frac{ n \rho- n_g \rho_z- c \delta t}{ c \tau_\mathrm{p}}\right]}{\rho} \bigg|_{x=-\frac{w}{2}}^{x=\frac{w}{2}} \bigg|_{x^\prime=-\frac{w}{2}- \delta x}^{x^\prime=- \delta x + \frac{w}{2}}     \right\}.
\end{multline}
Here, $\Lambda[x] \equiv \mathrm{max}\{ 1- |x|, 0 \}$ is the triangular function and we have used 
\begin{align}
\int_{-L/2}^{L/2} \dif z^\prime \int_{-L/2-z^\prime}^{L/2-z^\prime} \dif \rho_z f(\rho_z) = L \int_{-\infty}^\infty \dif \rho_z  \, \Lambda\left[ \frac{\rho_z}{L}    \right] f(\rho_z).
\end{align} 
 To simplify the first line of Eq.~\eqref{eq:GSR3}, we introduce spherical coordinates and use the delta distribution to carry out the $\rho$ integral. This way, Eq.~\eqref{eq:GSR3} reduces to
 \begin{multline} \label{eq:GSR4}
G_\mathrm{s} = -\frac{ \hbar}{8  L w^2 \tau_\mathrm{p}^2 \pi \epsilon_0 c n } \sum_{i = 1}^{4} (-1)^{i} \left\{  \int_{-1}^1 \dif \mathrm{cos}[\theta] \int_0^{2\pi} \dif \phi \frac{\rho}{|1 - \frac{n_g}{n} \mathrm{cos}(\theta)|}    \Lambda \left[ \frac{\rho_x - \delta x}{w}  \right] \Lambda\left[ \frac{\rho_y - \delta y}{w}  \right] \Lambda \left[ \frac{\rho_z}{L}\right]  \bigg|_{\rho = \rho_\mathrm{max} } \right. \\
\left.   - \frac{\tau_\mathrm{p}  c_n}{w}  \int \dif \rho_z \int \dif \rho_y  \Lambda \left[ \frac{\rho_z}{L}\right]   \frac{ \Lambda\left[ \frac{\rho_y - \delta y}{w}  \right]  \Lambda \left[\frac{n \rho- n_g \rho_z- c\delta t}{c\tau_\mathrm{p}}\right]}{\rho} \bigg|_{\rho_x= \rho_x^{(i)}}     \right\}.
\end{multline}
Here, we have introduced $\rho_\mathrm{max} = \mathrm{max} \{ \frac{c_n}{1 - \frac{n_g}{n} \mathrm{cos}(\theta)} \bar{\tau}_i, 0 \}$, $\bar{\tau}_2 = \bar{\tau}_4 = \delta t$, $\bar{\tau}_3 = \delta t - \tau_\mathrm{p}$ and $\bar{\tau}_4 = \delta t  + \tau_\mathrm{p}$, as well as $\rho_x^{(2)} = \rho_x^{(4)} = \delta x$, $\rho_x^{(1)} = \delta x + w $, and $\rho_x^{(3)} =  \delta x - w  $. \\
We can simplify Eq.~\eqref{eq:GSR4} further, by setting $\delta x = 0$ and assuming $ \delta y = \delta r   \gg w$. In this limit, the second row vanishes while the first becomes
 \begin{multline} \label{eq:GSRSimp}
G_\mathrm{s} =- \frac{ \hbar}{8  L w \tau_\mathrm{p}^2 \pi \epsilon_0 c n } \sum_{i = 1}^{4} (-1)^{i}  \int_{-1}^1 \dif \mathrm{cos}[\theta] \frac{\rho}{\mathrm{sin}(\theta)|1 - \frac{n_g}{n} \mathrm{cos}(\theta)|}    \Lambda\left[ \frac{\delta y - \rho \mathrm{sin}(\theta)}{w}  \right] \Lambda \left[ \frac{\rho \mathrm{cos}(\theta)}{L}\right]  \bigg|_{\rho = \mathrm{max} \{ \frac{c_n}{1 - \frac{n_g}{n} \mathrm{cos}(\theta)} \bar{\tau}_i, 0 \}} .
\end{multline}
This expression was numerically integrated to obtain the solid green line in Fig.~\ref{fig:EOSSignalDelt} of the main text.

\paragraph{Vacuum-Field Contribution.}

We use the expression for the correlation function in Eq.~\eqref{eq:CorrelationNormalMode}, which is valid for a real, constant refractive index, and insert it into Eq.~\eqref{eq:GVFFinal} to find the signal from vacuum-field fluctuations
 \begin{align} \label{eq:GVFNM1}
G_\mathrm{vac} = \frac{\mu_0 \hbar}{8\pi^2}\int \dif ^3 r^\prime\int \dif ^3 r^{\prime\prime} \int_{-\infty}^{\infty} \dif t  \int_{-\infty}^{\infty} \dif t^\prime L_{1}(\vec{r}^\prime, t) L_{2}(\vec{r}^{\prime\prime}, t^{\prime}) \left(  \frac{\partial^2}{\partial t \partial t^\prime} - c_n^2 \frac{\partial^2}{\partial x \partial x^\prime}\right) \frac{1}{\rho} \left(\frac{\mathcal{P}}{\frac{\rho}{c_n}- \tau} + \frac{\mathcal{P}}{\frac{\rho}{c_n}+ \tau}\right)  .
\end{align}
As in the last paragraph we integrate by parts to shift the derivatives to the pulse envelopes and introduce the coordinates $\tau$ and $\Greektens{$\rho$}$:
 \begin{multline} \label{eq:GVFNM2}
G_\mathrm{vac}  = \frac{ \hbar}{8  L w^2 \tau_\mathrm{p}^2 \pi^2 \epsilon_0 c n } \sum_{i = 1}^4 (-1)^i  \left\{  \int \dif^3 \rho    \Lambda \left[ \frac{\rho_x - \delta x}{w}  \right] \Lambda\left[ \frac{\rho_y- \delta y}{w}  \right] \Lambda\left[ \frac{\rho_z}{L}\right]   \frac{1}{\rho} \left[  \frac{\mathcal{P}}{\rho -c_n \tau} +\frac{\mathcal{P}}{\rho +c_n \tau}      \right] \bigg|_{\tau = n_g \frac{\rho_z}{c} +  \bar{\tau}_i} \right. \\
 \left. - \frac{\tau_\mathrm{p}  c_n}{w}  \int \dif \rho_z \int \dif \rho_y \int \dif \tau  \Lambda \left[ \frac{\rho_z}{L}\right]  \frac{ \Lambda\left[ \frac{\rho_y - \delta y}{w}  \right]  \Lambda \left[\frac{c \tau - n_g \rho_z- c \delta t}{ c \tau_\mathrm{p}}\right]}{\rho}  \left[  \frac{\mathcal{P}}{\rho -c_n \tau} +\frac{\mathcal{P}}{\rho +c_n \tau}      \right]\bigg|_{\rho_x = \rho_x^{(i)}}     \right\}.
\end{multline}
As before, we use $\delta x = 0 $ and set $\delta y = \delta r \gg w$ and find that the second row of Eq.~\eqref{eq:GVFNM2} vanishes and the first reduces to 
 \begin{multline} \label{eq:GVFNM3}
G_\mathrm{vac}  = \frac{ \hbar}{8  L w \tau_\mathrm{p}^2 \pi^2 \epsilon_0 c n } \sum_{i = 1}^4 (-1)^i  \int_0^\infty \dif \rho \int_{-1}^1 \dif \mathrm{cos}(\theta)    \Lambda \left[ \frac{\rho \mathrm{sin}(\theta)  - \delta r}{w}  \right]   \Lambda\left[ \frac{\rho \mathrm{cos}(\theta)}{L}\right]   \frac{1}{\mathrm{sin}(\theta)} \\
\times \left[  \frac{\mathcal{P}}{\rho -c_n \tau} +\frac{\mathcal{P}}{\rho +c_n \tau}      \right] \bigg|_{\tau = n_g \frac{\rho_z}{c}  + \bar{\tau}_i} 
\end{multline}
This expression was numerically integrated to obtain the solid red line in Fig.~\ref{fig:EOSSignalDelt} of the main text.

\subsection{Gaussian Pulses} \label{app:Gauss}

Apart from rectangular pulses, we also consider Gaussian pulses which have been used in the experiments in Refs.~\cite{benea-chelmus_electric_2019,settembrini2022detection}. They have a Gaussian shape in the $xy$-plane given by
\begin{align}\label{eq:GaussianTransverseMode}
g_{1}(\vec{r}_\parallel) & = g_{2}(\vec{r}_\parallel- \delta \vec{r}_\parallel)  = \sqrt{\frac{2}{\pi}} \frac{1}{w} \me^{- r_\parallel^2/w^2}.
\end{align}
Also the normalized spectrum is assumed to be Gaussian and given by
\begin{align} \label{eq:GaussianSpectrum}
\Eclo(\omega) & =\Eclt (\omega) \me^{\mi \omega \delta t} =\left( \frac{\tau_\sigma^2}{  2 \pi } \right)^{1/4} \me^{-\tau_\sigma^2 (|\omega | -\omega_c)^2/4},
\end{align} 
leading to the following spectral autocorrelation function
\begin{align}
f^2(\Omega) = \me^{-\Omega^2 \tau_\sigma^2 /4}  =  \me^{-\Omega^2 \tau_p^2 /8 \mathrm{ln}(2)} .
\end{align}
$\tau_\sigma$ is connected to the full width half maximum of the Gaussian pulse $\tau_p$ via $\tau_p = \tau_\sigma/2 \mathrm{ln}(2)$. Equations \eqref{eq:GaussianTransverseMode} and \eqref{eq:GaussianSpectrum} can be used to find space-time envelope of the laser pulses
\begin{align} \label{eq:laserPulseTimDomainGauss2}
L_1 & = \left( \frac{2}{\pi} \right)^{3/2} \frac{1}{\tau_\sigma w^2 L} \me^{-2 (n_g\frac{z}{c}- t)^2/\tau_\sigma^2} \me^{-2 r_\parallel^2/w^2} .
\end{align}
$L_2$ can be obtained via $L_2(\vec{r}, t) = L_1(\vec{r} + \delta \vec{r}_\parallel, t+ \delta t)$. \\

\paragraph{Vacuum Fluctuations.}

The signal from vacuum fluctuations in case of Gaussian pulses has been considered before \cite{lindel2020theory,lindel2022probing} and is given by
\begin{align} \label{eq:GVFGaussian}
G_\mathrm{vac}   = \frac{ \mu_0 \hbar  }{8 \pi^3 } \int_{0}^\infty \dif \Omega \, \Omega^2  \me^{-\Omega^2 \tau_\sigma^2 /4}  \int \dif^2 k_\parallel \,  \me^{-k_\parallel^2 w^2 /4} \me^{- \mi \delta \vec{r}_\parallel \cdot \vec{k}_\parallel } \left(1- \frac{k_x^2}{k^2}\right)  \mathrm{Re}[\Pi(k_z, \Omega)]    \cos [\Omega \delta t] ,
\end{align}
where we have defined the phase-matching function
\begin{align}
\Pi (k_z, \Omega) = \frac{1}{L k_z}\left[ \frac{\mi}{\frac{\Omega n_g}{c} + k_z}  + \frac{1- \me^{\mi L \left( \frac{\Omega n_g}{c} + k_z\right)}}{L \left(\frac{\Omega n_g}{c} + k_z \right)^2}      \right] + (n_g  \to -n_g ) .
\end{align}
Here, $+ (n_g  \to -n_g ) $ denotes adding the preceding expression subject to the replacement $n_g \to -n_g$. Equation \eqref{eq:GVFGaussian} can also be obtained from Eq.~\eqref{eq:GVFFreq} without any further approximations. It is valid for a general complex refractive index in the THz frequency range $n(\Omega)$ and thus includes absorption and dispersion effects. Equation \eqref{eq:GVFGaussian} was numerically integrated to obtain the dashed green lines in Fig.~\ref{fig:EOSSignalDelt} and \ref{fig:TFDT} of the main text.


\paragraph{Source Radiation.}

To obtain the signal stemming from source radiation, we insert the Green tensor in Eq.~\eqref{eq:GreensGeneral} into Eq.~\eqref{eq:GRSRSimplified} and find
\begin{multline} \label{eq:GRSRSimplifiedGaussian}
G_{\theta_1\theta_2} \big |_\mathrm{s}   = \frac{ \mu_0 \hbar  }{16 \pi^3  L^2 }\int_{V_C} \dif^3 r^\prime  g^{(1)2}(\vec{r}_\parallel^\prime) \int_{V_C} \dif^3 r^{\prime\prime} g^{(2)2}(\vec{r}_\parallel^{\prime\prime}) \int_{-\infty}^\infty \dif \Omega \, \Omega^2 f^2(\Omega)  \int \dif^2 k_\parallel \me^{\mi \vec{k}_\parallel \cdot(\vec{r}_\parallel^\prime- \vec{r}_\parallel^{\prime\prime} )} \left(1- \frac{k_x^2}{k^2}\right) \\
 \times   \left\{ \mathrm{Im}\left[  P(\theta_1) P(\theta_2)  \right]  \mathrm{Im} \left[ \frac{\me^{\mi k_z |z^\prime-z^{\prime\prime}|}}{k_z} \right]    \cos [\Omega \phi]   +\mathrm{Im}\left[  P(\theta_1) P^\ast(\theta_2)  \right]   \mathrm{Re}\left[\frac{\me^{\mi k_z |z^\prime-z^{\prime\prime}|}}{k_z} \right]    \sin [\Omega \phi] \right\} .
\end{multline}
Introducing polar coordinates $k_x = k_\parallel \mathrm{sin}(\phi)$ and $k_y = k_\parallel \mathrm{cos}(\phi)$, and performing the $\vec{r}_\parallel$, $z$, and $\phi $ integrals, we find
\begin{multline} \label{eq:GRSRSimplifiedGaussian2}
G_{\theta_1\theta_2} \big |_\mathrm{s}  = \frac{ \mu_0 \hbar  }{16 \pi^3 } \int_{0}^\infty \dif \Omega \, \Omega^2  \me^{-\Omega^2 \tau_\sigma^2 /4}  \int_0^\infty \dif k_\parallel \, k_\parallel  \me^{-k_\parallel^2 w^2 /4} \beta(k_\parallel, \delta \vec{r}_\parallel) \\
 \times   \left\{ \mathrm{Im}\left[  P(\theta_1) P(\theta_2)  \right]  \mathrm{Im}[ \Pi(k_z, \Omega)]    \cos [\Omega \delta t ]  -\mathrm{Im}\left[  P(\theta_1) P^\ast(\theta_2)  \right]   \mathrm{Re}[\Pi(k_z, \Omega)]    \sin [\Omega \delta t] \right\} .
\end{multline}
Here, we have defined
\begin{align}
\frac{1}{2\pi}\beta(k_\parallel, \delta \vec{r}_\parallel) = \left\{ \begin{array}{ll}
  J_0[k_\parallel \delta y] - \frac{k_\parallel}{k^2 \delta y}J_1[k_\parallel \delta y] & \mathrm{if }\,\delta\vec{r}_\parallel = \delta y \vec{e}_y \\
\left(1- \frac{k_\parallel}{k^2 \delta x} \right)J_0[k_\parallel \delta x] + \frac{k_\parallel}{k^2 \delta x} J_1[k_\parallel \delta x] & \mathrm{if } \,\delta\vec{r}_\parallel = \delta x \vec{e}_x 
\end{array} \right. ,
\end{align}
where $J_n$ denotes the Bessel functions of the first kind. If $P(\theta_1 ) = \mi $ and $P(\theta_2) = 1$ \footnote{This means that source radiation propagating from mode $2$ to mode $1$ is probed. If $\delta t > 0 $ mode $1$ is delayed with respect to mode two.}, we find
\begin{align} \label{eq:GRSRSimplifiedGaussian3}
G_\mathrm{s}   = \frac{ \mu_0 \hbar  }{16 \pi^3 } \int_{0}^\infty \dif \Omega \, \Omega^2  \me^{-\Omega^2 \tau_\sigma^2 /4}  \int_0^\infty \dif k_\parallel \,k_\parallel  \me^{-k_\parallel^2 w^2 /4}  \beta(k_\parallel, \delta \vec{r}_\parallel)\left(  \mathrm{Im}[ \Pi(k_z, \Omega)]    \cos [\Omega \delta t ]  -   \mathrm{Re}[\Pi(k_z, \Omega)]    \sin [\Omega \delta t] \right) .
\end{align}
This expression was numerically integrated with $\delta \vec{r}_\parallel = \delta r \vec{e}_y$ to obtain the dashed red curve in Fig.~\ref{fig:EOSSignalDelt} in the main text. The signal stemming from the dissipative part of the response function $G_{\mathcal{R}^{\prime\prime}} $ shown in Fig.~\ref{fig:TFDT} can be obtained similarly from Eq.~\eqref{eq:GRSRSimplifiedGaussian2}:
\begin{align} \label{eq:GRSRSimplifiedGaussian4}
G_{\mathcal{R}^{\prime\prime}}  = -\frac{ \mu_0 \hbar  }{16 \pi^3 } \int_{0}^\infty \dif \Omega \, \Omega^2  \me^{-\Omega^2 \tau_\sigma^2 /4}  \int_0^\infty \dif k_\parallel \,k_\parallel  \me^{-k_\parallel^2 w^2 /4}  \beta(k_\parallel, \delta \vec{r}_\parallel)     \mathrm{Re}[\Pi(k_z, \Omega)]    \sin [\Omega \delta t]  .
\end{align}

\section{Refractive Index} \label{app:Param}

The refractive index used to simulate the EOS signal displayed in Figs.~\ref{fig:EOSSignalDelt} (only dashed lines) and \ref{fig:TFDT} in the main text, is the one measured in Ref.~\cite{leitenstorfer1999detectors} for GaP. The group refractive index for the NIR laser pulses is given by $n_g = 3.556$ at a central wave-length of the laser pulse of $835\,\mathrm{nm}$. In the THz we use the complex valued, dispersive refractive index for GaP given in Ref.~\cite{leitenstorfer1999detectors}. For the solid lines in Fig.~\ref{fig:EOSSignalDelt} in the main text we used $n_g = 3.556$ and $n(\Omega) \approx n = 3.33$ as also stated in the main text.

\end{widetext}


%

\end{document}